\documentclass{aa}  

\usepackage{graphicx}
\usepackage{amsmath,amsfonts,amssymb}
\usepackage{natbib}

\usepackage{txfonts}
\usepackage{xcolor}

\usepackage{blindtext}
\usepackage{float}
\usepackage{dblfloatfix}
\usepackage{afterpage}
\usepackage{ifthen}
\usepackage[morefloats=12]{morefloats}

\usepackage{placeins}
\usepackage{multicol}
\bibpunct{(}{)}{;}{a}{}{,}
\usepackage[switch]{lineno}
\definecolor{linkcolor}{rgb}{0.6,0,0}
\definecolor{citecolor}{rgb}{0,0,0.75}
\definecolor{urlcolor}{rgb}{0.12,0.46,0.7}
\usepackage[breaklinks, colorlinks, urlcolor=urlcolor,
    linkcolor=linkcolor,citecolor=citecolor,pdfencoding=auto]{hyperref}
\hypersetup{linktocpage}
\usepackage{bold-extra}
\usepackage{makecell}

\usepackage[capitalise]{cleveref}
\Crefname{section}{Sect.}{Sects.}
\Crefname{table}{Table}{Tables}
\Crefname{equation}{Eq.}{Eqs.}
\Crefname{appendix}{Appendix}{Appendices}
\makeatletter
\AddToHook{cmd/appendix/before}{\def\cref@section@alias{appendix}}
\makeatother

\def\setsymbol#1#2{\expandafter\def\csname #1\endcsname{#2}}
\def\getsymbol#1{\csname #1\endcsname}

\def\Planck{\textit{Planck}}

\newbox\tablebox    \newdimen\tablewidth
\def\leaderfil{\leaders\hbox to 5pt{\hss.\hss}\hfil}
\def\endPlancktablewide{\tablewidth=\textwidth 
    $$\hss\copy\tablebox\hss$$
    \vskip-\lastskip\vskip -2pt}
\def\tablenote#1 #2\par{\begingroup \parindent=0.8em
    \abovedisplayshortskip=0pt\belowdisplayshortskip=0pt
    \noindent
    $$\hss\vbox{\hsize\tablewidth \hangindent=\parindent \hangafter=1 \noindent
    \hbox to \parindent{$^#1$\hss}\strut#2\strut\par}\hss$$
    \endgroup}
\def\doubleline{\vskip 3pt\hrule \vskip 1.5pt \hrule \vskip 5pt}

\def\L2{\ifmmode L_2\else $L_2$\fi}

\def\DeltaT{\ifmmode \Delta T\else $\Delta T$\fi}
\def\deltat{\ifmmode \Delta t\else $\Delta t$\fi}
\def\fknee{\ifmmode f_{\rm knee}\else $f_{\rm knee}$\fi}
\def\Fmax{\ifmmode F_{\rm max}\else $F_{\rm max}$\fi}
\def\solar{\ifmmode{\rm M}_{\mathord\odot}\else${\rm M}_{\mathord\odot}$\fi}
\def\Msolar{\ifmmode{\rm M}_{\mathord\odot}\else${\rm M}_{\mathord\odot}$\fi}
\def\Lsolar{\ifmmode{\rm L}_{\mathord\odot}\else${\rm L}_{\mathord\odot}$\fi}
\def\inv{\ifmmode^{-1}\else$^{-1}$\fi}
\def\mo{\ifmmode^{-1}\else$^{-1}$\fi}
\def\sup#1{\ifmmode ^{\rm #1}\else $^{\rm #1}$\fi}
\def\expo#1{\ifmmode \times 10^{#1}\else $\times 10^{#1}$\fi}
\def\,{\thinspace}
\def\lsim{\mathrel{\raise .4ex\hbox{\rlap{$<$}\lower 1.2ex\hbox{$\sim$}}}}
\def\gsim{\mathrel{\raise .4ex\hbox{\rlap{$>$}\lower 1.2ex\hbox{$\sim$}}}}

\def\simprop{\mathrel{\raise .4ex\hbox{\rlap{$\propto$}\lower 1.2ex\hbox{$\sim$}}}}
\def\deg{\ifmmode^\circ\else$^\circ$\fi}
\def\pdeg{\ifmmode $\setbox0=\hbox{$^{\circ}$}\rlap{\hskip.11\wd0 .}$^{\circ}
          \else \setbox0=\hbox{$^{\circ}$}\rlap{\hskip.11\wd0 .}$^{\circ}$\fi}
\def\arcs{\ifmmode {^{\scriptstyle\prime\prime}}
          \else $^{\scriptstyle\prime\prime}$\fi}
\def\arcm{\ifmmode {^{\scriptstyle\prime}}
          \else $^{\scriptstyle\prime}$\fi}
\newdimen\sa  \newdimen\sb
\def\parcs{\sa=.07em \sb=.03em
     \ifmmode \hbox{\rlap{.}}^{\scriptstyle\prime\kern -\sb\prime}\hbox{\kern -\sa}
     \else \rlap{.}$^{\scriptstyle\prime\kern -\sb\prime}$\kern -\sa\fi}
\def\parcm{\sa=.08em \sb=.03em
     \ifmmode \hbox{\rlap{.}\kern\sa}^{\scriptstyle\prime}\hbox{\kern-\sb}
     \else \rlap{.}\kern\sa$^{\scriptstyle\prime}$\kern-\sb\fi}
\def\ra[#1 #2 #3.#4]{#1\sup{h}#2\sup{m}#3\sup{s}\llap.#4}
\def\dec[#1 #2 #3.#4]{#1\deg#2\arcm#3\arcs\llap.#4}
\def\deco[#1 #2 #3]{#1\deg#2\arcm#3\arcs}
\def\rra[#1 #2]{#1\sup{h}#2\sup{m}}

\def\dots{\relax\ifmmode \ldots\else $\ldots$\fi}
\def\WHzsr{\ifmmode $W\,Hz\mo\,sr\mo$\else W\,Hz\mo\,sr\mo\fi}
\def\mHz{\ifmmode $\,mHz$\else \,mHz\fi}
\def\GHz{\ifmmode $\,GHz$\else \,GHz\fi}
\def\mKs{\ifmmode $\,mK\,s$^{1/2}\else \,mK\,s$^{1/2}$\fi}
\def\muKs{\ifmmode \,\mu$K\,s$^{1/2}\else \,$\mu$K\,s$^{1/2}$\fi}
\def\muKRJs{\ifmmode \,\mu$K$_{\rm RJ}$\,s$^{1/2}\else \,$\mu$K$_{\rm RJ}$\,s$^{1/2}$\fi}
\def\muKHz{\ifmmode \,\mu$K\,Hz$^{-1/2}\else \,$\mu$K\,Hz$^{-1/2}$\fi}
\def\MJysr{\ifmmode \,$MJy\,sr\mo$\else \,MJy\,sr\mo\fi}
\def\MJysrmK{\ifmmode \,$MJy\,sr\mo$\,mK$_{\rm CMB}\mo\else \,MJy\,sr\mo\,mK$_{\rm CMB}\mo$\fi}
\def\microns{\ifmmode \,\mu$m$\else \,$\mu$m\fi}

\def\muK{\ifmmode \,\mu$K$\else \,$\mu$\hbox{K}\fi}
\def\microK{\ifmmode \,\mu$K$\else \,$\mu$\hbox{K}\fi}
\def\muW{\ifmmode \,\mu$W$\else \,$\mu$\hbox{W}\fi}
\def\kms{\ifmmode $\,km\,s$^{-1}\else \,km\,s$^{-1}$\fi}
\def\kmsMpc{\ifmmode $\,\kms\,Mpc\mo$\else \,\kms\,Mpc\mo\fi}
\providecommand{\sorthelp}[1]{}

\newcommand{\mathsc}[1]{{\normalfont\textsc{#1}}}
\newcommand{\dv}[0]{\vec{d}}
\newcommand{\s}[0]{\vec{s}}
\renewcommand{\r}[0]{\vec{r}}
\newcommand{\M}[0]{\tens{M}}
\renewcommand{\P}[0]{\tens{P}}
\newcommand{\G}[0]{\tens{G}}
\newcommand{\B}[0]{\tens{B}}
\renewcommand{\a}[0]{\vec{a}}
\newcommand{\n}[0]{\vec{n}}
\renewcommand{\t}[0]{\vec{t}}
\def\Cosmoglobe{\textsc{Cosmoglobe}}
\def\Planck{\textit{Planck}}

\def\COBE{\textit{COBE}}
\def\GAIA{\textit{Gaia}}

\def\Gaia{\textit{Gaia}}
\def\WISE{WISE}

\def\IRAS{\textit{{IRAS}}}
\def\nside{$N_{\mathrm{side}}$}

\newcommand{\CII}{\ensuremath{\mathsc{C\ ii}}}

\newcommand{\HI}{\ensuremath{\mathsc{H\ i}}}
\def\Commander{\texttt{Commander} }

\def\Tcmb{\ifmmode T_\mathrm{CMB}\else $T_{\mathrm{CMB}}$\fi}
\def\Tcold{\ifmmode T_\mathrm{c}\else $T_{\mathrm{c}}$\fi}
\def\Thot{\ifmmode T_\mathrm{h}\else $T_{\mathrm{h}}$\fi}
\def\Tnear{\ifmmode T_\mathrm{n}\else $T_{\mathrm{n}}$\fi}
\def\Thalpha{\ifmmode T_\mathrm{H\alpha}\else $T_{\mathrm{H\alpha}}$\fi}
\def\scmb{\ifmmode s_\mathrm{CMB}\else $s_{\mathrm{CMB}}$\fi}
\def\squad{\ifmmode s_\mathrm{quad}\else $s_{\mathrm{quad}}$\fi}
\def\ssynch{\ifmmode s_\mathrm{s}\else $s_\mathrm{s}$\fi}
\def\sdust{\ifmmode s_\mathrm{d}\else $s_{\mathrm{d}}$\fi}
\def\ssdust{\ifmmode s_\mathrm{sd}\else $s_{\mathrm{sd}}$\fi}
\def\same{\ifmmode s_\mathrm{AME}\else $s_{\mathrm{AME}}$\fi}
\def\ssrc{\ifmmode s_\mathrm{src}\else $s_{\mathrm{src}}$\fi}
\def\sco{\ifmmode s_\mathrm{CO}\else $s_{\mathrm{CO}}$\fi}
\def\sff{\ifmmode s_\mathrm{ff}\else $s_{\mathrm{ff}}$\fi}
\def\gff{\ifmmode g_\mathrm{ff}\else $g_{\mathrm{ff}}$\fi}
\def\fsynch{\ifmmode f_\mathrm{s}\else $f_{\mathrm{s}}$\fi}
\def\fsd{\ifmmode f_\mathrm{sd}\else $f_{\mathrm{sd}}$\fi}
\def\fame{\ifmmode f_\mathrm{AME}\else $f_{\mathrm{AME}}$\fi}
\def\alphasrc{\ifmmode \alpha_\mathrm{src}\else $\alpha_{\mathrm{src}}$\fi}
\def\bcold{\ifmmode \beta_\mathrm{c}\else $\beta_{\mathrm{c}}$\fi}
\def\bhot{\ifmmode \beta_\mathrm{h}\else $\beta_{\mathrm{h}}$\fi}
\def\bnear{\ifmmode \beta_\mathrm{n}\else $\beta_{\mathrm{n}}$\fi}
\def\bhalpha{\ifmmode \beta_\mathrm{H\alpha}\else $\beta_{\mathrm{H\alpha}}$\fi}
\def\bsynch{\ifmmode \beta_\mathrm{s}\else $\beta_{\mathrm{s}}$\fi} 
\def\bsun{\ifmmode \beta_\mathrm{sun}\else $\beta_{\mathrm{sun}}$\fi} 
\def\nuzeros{\ifmmode \nu_{0,\mathrm{s}}\else $\nu_{0,\mathrm{s}}$\fi} 
\def\nuzeroff{\ifmmode \nu_{0,\mathrm{ff}}\else $\nu_{0,\mathrm{ff}}$\fi} 
\def\nuzerocold{\ifmmode \nu_{0,\mathrm{c}}\else $\nu_{0,\mathrm{c}}$\fi}
\def\nuzerohot{\ifmmode \nu_{0,\mathrm{h}}\else $\nu_{0,\mathrm{h}}$\fi}
\def\nuzeronear{\ifmmode \nu_{0,\mathrm{n}}\else $\nu_{0,\mathrm{n}}$\fi} 
\def\nuzerohalpha{\ifmmode \nu_{0,\mathrm{H\alpha}}\else$\nu_{0,\mathrm{H\alpha}}$\fi} 
\def\nuzeroame{\ifmmode \nu_{0,\mathrm{AME}}\else $\nu_{0,\mathrm{AME}}$\fi} 
\def\nuzerosd{\ifmmode \nu_{0,\mathrm{}}\else $\nu_{0,\mathrm{sd}}$\fi} 
\def\nuzerosrc{\ifmmode \nu_{0,\mathrm{src}}\else $\nu_{0,\mathrm{src}}$\fi} 
\def\nup{\ifmmode \nu_{\mathrm{p}}\else $\nu_{\mathrm{p}}$\fi} 
\def\alphasd{\ifmmode \alpha_{\mathrm{sd}}\else $\alpha_{\mathrm{sd}}$\fi} 
\def\Te{\ifmmode T_{\mathrm{e}}\else $T_{\mathrm{e}}$\fi} 
\def\kB{\ifmmode k_\mathrm{B}\else $k_{\mathrm{B}}$\fi} 

\newcommand{\x}{\checkmark}

\begin{document}

%\title{\bfseries{\Cosmoglobe\ DR2. VII. Multi-component spectral energy density model of large-scale thermal dust emission from 100\,GHz to 100\,THz}}
\title{\bfseries{\Cosmoglobe\ DR2. VII. Towards a concordance model of large-scale thermal dust emission for microwave and infrared frequencies}}
%\title{\bfseries{\Cosmoglobe\ DR2. VII. The mean thermal dust emission SED as seen by \Planck\ HFI and \COBE-DIRBE}}

   \newcommand{\oslo}[0]{1}
\newcommand{\milan}[0]{2}
\newcommand{\ijclab}[0]{3}
\newcommand{\gothenberg}[0]{4}
\newcommand{\trento}[0]{5}
\newcommand{\milanoinfn}[0]{6}
\author{\small
E.~Gjerl\o w\inst{\oslo}\thanks{Corresponding author: E.~Gjerløw; \url{eirik.gjerlow@astro.uio.no}}
\and
R.~M.~Sullivan\inst{\oslo}
\and
R.~Aurvik\inst{\oslo}
\and
A.~Basyrov\inst{\oslo}
\and
L.~A.~Bianchi\inst{\oslo}
\and
A.~Bonato\inst{\milan}
\and
M.~Brilenkov\inst{\oslo}
\and
H.~K.~Eriksen\inst{\oslo}
\and
U.~Fuskeland\inst{\oslo}
\and
M.~Galloway\inst{\oslo}
\and
K.~A.~Glasscock\inst{\oslo}
\and
L.~T.~Hergt\inst{\ijclab}
\and
D.~Herman\inst{\oslo}
\and
J.~G.~S.~Lunde\inst{\oslo}
\and
A.~I.~Silva Martins\inst{\oslo}
\and
M.~San\inst{\oslo}
\and
D.~Sponseller\inst{\gothenberg}
\and
N.-O.~Stutzer\inst{\oslo}
\and
H.~Thommesen\inst{\oslo}
\and
V.~Vikenes\inst{\oslo}
\and
D.~J.~Watts\inst{\oslo}
\and
I.~K.~Wehus\inst{\oslo}
\and
L.~Zapelli\inst{\milan,\trento,\milanoinfn}
}
\institute{\small
Institute of Theoretical Astrophysics, University of Oslo, Blindern, Oslo, Norway\goodbreak
\and
Dipartimento di Fisica, Università degli Studi di Milano, Via Celoria, 16, Milano, Italy
\and
Laboratoire de Physique des 2 infinis -- Irène Joliot Curie (IJCLab), Orsay, France
\and
Department of Space, Earth and Environment, Chalmers University of Technology, Gothenburg, Sweden\goodbreak
\and
Università di Trento, Università degli Studi di Milano, CUP E66E23000110001\goodbreak
\and
INFN sezione di Milano, 20133 Milano, Italy\goodbreak
}

   %\institute{Institute of Theoretical Astrophysics, University of Oslo, Blindern, Oslo, Norway}
  
   % Shortened title, author list for top of page 
   \titlerunning{Towards a concordance model for thermal dust emission}
   \authorrunning{Gjerløw et al.}

   \date{\today} 
   
   \abstract{
     We fit a four-component thermal dust model to \COBE-DIRBE data between 3.5 and 240\,$\mu$m within the global Bayesian end-to-end \Cosmoglobe\ DR2 reanalysis. Following a companion analysis of \Planck\ HFI, the four components of this model correspond to ``hot dust'', ``cold dust'', ``nearby dust'', and ``H$\alpha$ correlated dust'', respectively, and each component is modelled in terms of a fixed spatial template and a spatially isotropic spectral energy density (SED) defined by an overall free amplitude for each DIRBE channel. Except for the cold dust amplitude, which is only robustly detected in the 240\,$\mu$m channel, we measure statistically significant template amplitudes for all components in all DIRBE channels between 12 and $240\,\mu\mathrm{m}$. In the 3.5 and $4.9\,\mu$m channels, only the hot component is detected, while the 1.25 and 2.2$\,\mu$m channels are too dominated by starlight emission to allow robust dust detections. The total number of DIRBE-specific degrees of freedom in this model is 25. Despite this low dimensionality, the resulting total SED agrees well with recent \texttt{astrodust} predictions. At both low and high frequencies, more than 95\,\% of the frequency map variance is captured by the model, while at 60 and $100\,\mu\mathrm{m}$ about 70\,\% of the signal variance is successfully accounted for. The hot dust component, which in a companion paper has been found to correlate strongly with \ion{C}{II} emission, has the highest absolute amplitude in all DIRBE frequency channels; in particular, at 3.5\,$\mu$m, which is known to be dominated by polycyclic aromatic hydrocarbon emission, this component accounts for at least 80\,\% of the total signal. This analysis represents an important step towards establishing a joint concordance model of thermal dust emission applicable to both the microwave and infrared regimes.
   }

   \maketitle

%\setcounter{tocdepth}{2}
%\tableofcontents
   
% INTRODUCTION
%-------------------------------------------------------------------
\section{Introduction}

In a series of seven companion papers, within which this is the last,
we have reanalyzed the 30-year-old \COBE-DIRBE data using modern
end-to-end Bayesian statistical techniques as implemented in the
\Cosmoglobe\footnote{\url{http://cosmoglobe.uio.no}} framework. From this work, a set of ten
full-sky DIRBE frequency maps has emerged \citep{CG02_01}, covering the infrared frequency range
between 1.25 and 240\,$\mu$m. These maps have both substantially lower
systematic errors and better error characterization compared to their
official counterparts \citep{hauser1998,CG02_01}, and, in particular,
they suffer far less from zodiacal light contamination
\citep{K98,CG02_02}. As a result they can be used for more detailed
astrophysics applications. A few examples of this are presented in the
current paper suite, including improved estimates of the cosmic infrared
background (CIB) spectrum \citep{CG02_03} and large-scale starlight emission
\citep{CG02_04}.

The main topic of the last three papers in the series (\citet{CG02_05,
  CG02_06}; and the present paper) is global modelling of thermal dust
radiation on large angular scales in the microwave and infrared
frequency regimes. This issue has been the focus of intense scrutiny
ever since the groundbreaking IRAS \citep{neugebauer:1984}
measurements were published in 1982, and its scientific importance has
only increased through the release of a series of increasingly
sensitive data sets, such as \COBE-DIRBE \citep{hauser1998} and
\Planck\ HFI \citep{planck2016-l03}. Today, detailed dust modelling
plays a key role in many of the most competitive fields of cosmology,
from the search for inflationary gravitational waves in cosmic
microwave background (CMB) $B$-mode polarization data
\citep{bicep2021} to measurements of dark energy using distant
supernovae \citep{popovic:2025}.

As of today, \Planck\ HFI defines the state-of-the-art for full-sky thermal
dust mapping, both in terms of signal-to-noise ratio as well as systematic
control. Based on \Planck's nine frequency channels, the team
produced several exquisite dust models in both intensity and
polarization
\citep{planck2013-XVII,planck2013-p06b,planck2013-XIV,planck2014-XIX,planck2014-a12,planck2014-XXII,planck2016-l11A,planck2016-l11B},
and these now form the basis for much of the dust modelling efforts in
the field \citep[e.g.,][]{pysm2,pysm3}. However, at the same time, the
relatively narrow frequency range of \Planck, covering only 30-857\,GHz,
implies that the applicability of these models is currently quite
limited. Furthermore, the absolute calibration of the 857\,GHz \Planck\
channel, which is nominally the most sensitive \Planck\ dust channel,
is uncertain at the $\sim$10\,\% level \citep{planck2016-l03,npipe},
and this induces a large uncertainty on the dust spectral energy density (SED) parameters when
extrapolating to higher frequencies.

In the current paper, we address these issues by fitting the
multi-component dust model proposed by \citet{CG02_05} and
\citet{CG02_06} to the re-processed \COBE-DIRBE data within the
\Cosmoglobe\ DR2 analysis framework. This model consists of four
primary components, namely 1) cold dust, 2) hot dust, 3) nearby dust,
and 4) H$\alpha$-correlated dust (which observed in extinction). From
the previous papers, this model is already known to fit the
\Planck\ HFI frequencies very well when coupled to simple modified
blackbody (MBB) SEDs with spatially constant spectral parameters, and
in this paper we show that the same spatial morphologies also trace
dust in the DIRBE frequencies with high precision, although with more
complicated SED behaviour. The combined result is a global model that
jointly describes both microwave and infrared frequencies.

So far, this model has only been developed for and applied to
intensity measurements. However, polarized thermal dust emission also
plays a key role in modern astrophysics and cosmology. For instance,
massive resources are currently being spent on searching for and
constraining the amplitude of inflationary gravitational waves through
deep CMB $B$-mode polarization experiments \citep{litebird2022, SO2019}, and polarized thermal dust
emission represents a key challenge for these \citep{fuskeland:2023}. Given the high
efficiency of the \Cosmoglobe\ DR2 four-component dust model for
intensity data, it is reasonable to expect a similar performance for
polarization observations. 

\FloatBarrier
\begin{figure*}[t]
  \centering
  \includegraphics[width=0.49\linewidth]{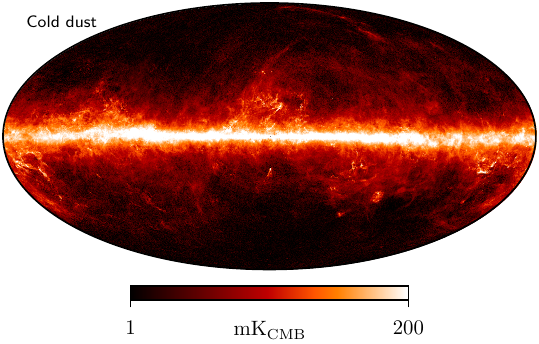}
  \includegraphics[width=0.49\linewidth]{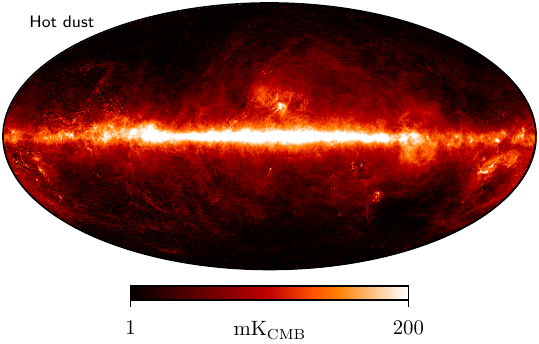}\\
  \includegraphics[width=0.49\textwidth]{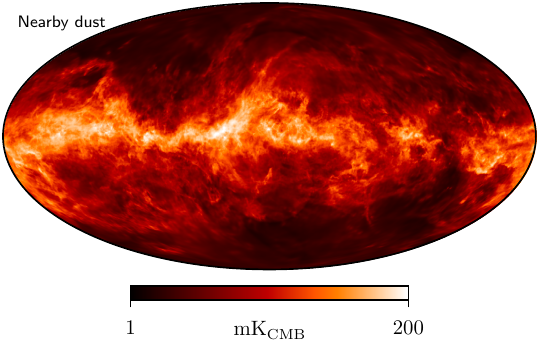}
  \includegraphics[width=0.49\textwidth]{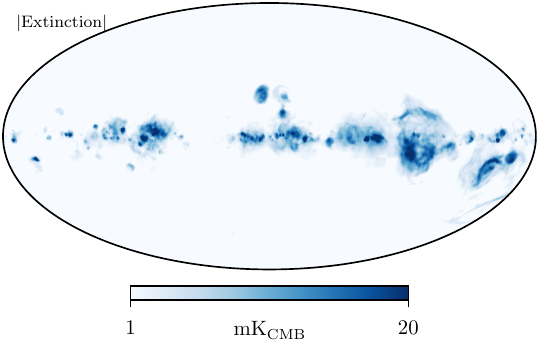}
  \caption{Dust template maps used in the \Cosmoglobe\ DR2 sky model, whose units are defined with the \Planck\ HFI 545-1 bolometer channel as the reference frequency. From left to right and top to bottom, the four panels show 1) the cold dust template, $\t_{\mathrm{cold}}$; 2) the hot dust template, $\t_{\mathrm{hot}}$; 3) the nearby dust template, $\t_{\mathrm{nearby}}$; and 4) the (absolute value of the) H$_{\alpha}$-correlated dust extinction template, $\a_{\mathrm{H}\alpha}$. All panels employ the \Planck\ non-linear high dynamic range color scheme, defined by $\log_{10}((\t + \sqrt{4+\t^2})/2)$, which results in a nearly linear behaviour for small values and exponential for large values.  }
  \label{fig:templates}
\end{figure*}

\section{Bayesian modelling of thermal dust emission in \Cosmoglobe\ DR2}
\label{sec:dr2}

The following work is part of \Cosmoglobe\ Data Release 2, and
represents as such a global Bayesian analysis exploration of the joint
posterior of all involved parameters. The following analysis is thus
intricately bound together with the other papers, both in terms of the
data sets and the data model used. Hence, we will in the following
only give a brief recap of the underlying data model and algorithms,
while paying special attention to the thermal dust model. For full
details, we refer the interested reader to \citet{CG02_01} and
references therein.

\subsection{Data model}
In the parametric Bayesian \Cosmoglobe\ framework
\citep[e.g.,][]{bp01,watts2023_dr1,CG02_01}, the first step is to
write down an explicit parametric model that incorporates all known
aspects of the dataset with which we are working, including both
instrumental and astrophysical effects. The main target of the current
analysis are the DIRBE time-ordered data (TOD), and we adopt the
following model for these,
\begin{align}
	\label{eq:model}
    \dv_{\mathrm{DIRBE}}(\nu)
    &=\G\P\B\sum_{c=1}^{n_{\mathrm{comp}}}\M_c(\nu)\a_c+\s_{\mathrm{zodi}}(\nu) +
          \s_{\mathrm{static}}(\nu) + \n \\
                         & \equiv \s^{\mathrm{tot}} +
    \n_{\mathrm{corr}} + \n_{\mathrm{w}} \nonumber.
\end{align}
Here $\nu$ is the frequency of a given data channel; $\dv$ is a vector
containing the full TOD; $\G$ is an overall gain factor; $\P$ is the
pointing matrix which projects the pixelated sky onto a
$n_{\mathrm{tod}}$-sized space; $\B$ is the instrumental beam
convolution operator; and $\n_{\mathrm{corr}}$ and $\n_{\mathrm{w}}$
represents instrumental correlated and white noise, respectively, with
spectral parameters $\xi_{\mathrm{n}}$, most importantly the white
noise rms per TOD sample.

The physical sky is represented by three terms. Firstly, the sum over
$c$ includes all stationary sky components, such as Galactic dust or
free-free emission. Each of these are modelled in terms of a mixing
matrix, $\M_c$, that describes the relative amplitude of the given
component as a function of frequency, and an amplitude map, $\a$.
Secondly, $\s_{\mathrm{zodi}}(\nu)$ describes zodiacal emission, i.e.,
emission from dust grains within the Solar system
\citep{K98,CG02_02}. Because these grains are located so close to the
Earth, this component must be modelled dynamically by taking into
account the observatory's position in the Solar system as a function
of time. Finally $\s_{\mathrm{static}}$ accounts for emission that
appears static in Solar-centric coordinates, which for instance could
be due to either errors in the zodiacal light model or DIRBE
sidelobes. These two last terms are discussed extensively by
\citet{CG02_01} and \citet{CG02_02}. In the present work, however, we
focus on the first of the three terms.

%All DIRBE
%frequency sky maps are pixelized using
%HEALPix\footnote{\url{http://healpix.sourceforge.net/}}
%\citep{healpix,Zonca2019} with a resolution parameter of
%$N_{\mathrm{side}}=512$.

%As it stands, this is a ``complete'' model for the frequency domain we are interested in this paper, but we expect that it will have to expand as additional data is added and more discoveries are made.

\subsection{Multi-component thermal dust modelling}
\label{sec:modelling}
%\section{Dust modelling}
%\subsection{Current status}
%\label{sec:current_dust}
Interstellar dust --- amorphous particles of silicate and carbonaceous
materials --- makes its presence known on practically all
astrophysically relevant wavelengths. Understanding this material is
interesting in its own right, however, it is also crucial to improve
astrophysical foreground removal, especially when interstellar dust
emission contaminates other signals of interest
\citep[e.g.,][]{Hensley2021}.

Recently, a major step forward in this direction was made by
\citet{Hensley2023}, who introduced the so-called ``astrodust+PAH''
model, wherein the diffuse interstellar medium (ISM) was hypothesized
to be made up of a single composite material (the eponymous astrodust)
for scales larger than $\sim0.02~\mu$m, and a distinct variety of
materials --- including so-called polycyclic aromatic hydrocarbons
(PAH) --- on scales smaller than this. In the wavelength regime
between $3000-100~\mu$m, this model is well described by an MBB
SED \footnote{The actual astrodust model is made up of a composite MBB
which has a transition between 353 and 217 GHz}, i.e., an SED that
follows
\begin{equation}
s(\nu) \propto \nu^\beta B(\nu, T),
\label{eq:mbb}
\end{equation}
where $\nu$ is the frequency, $B$ is the Planck law for a perfect
blackbody with temperature $T$, and $\beta$ is the spectral index.
Typical ISM temperatures are around $\sim$20\,K, for which the SED
typically peaks around 150\,$\mu$m or $\sim$2000\,GHz. At shorter
wavelengths (2.5\,$\mu$m to 12\,$\mu$m), the astrodust+PAH model is
dominated by nanoscale particle emission, and exhibits strong emission
lines at various wavelengths (see, e.g., Fig.~10 in \citealt{Hensley2023}).

\subsubsection{The four-component \Cosmoglobe\ DR2 dust model}
\label{sec:fourcompmodel}
The astrodust model provides a physically well understood framework
for modelling Galactic thermal dust emission. Typically, in a given
line-of-sight, the relative contribution of various dust components
will vary. At the same time, the degree to which such variations can
be detected and described is limited by the resolution and
signal-to-noise ratios of the available data at the wavelengths
involved. Thus, classifying populations of interstellar dust with
common spectral parameters has been of high importance.

It was demonstrated by \citet{CG02_05} that a natural and highly
effective classification of such populations can be achieved through
the use of templates derived from surveys of spectral line emission
(\CII, H$\alpha$, CO and \HI) and from inference of nearby dust
structures via starlight extincion \citep{edenhofer:2024}. In that
paper, we showed that a linear combination of five such templates
could explain more than 95\% of the large-scale dust signal variance
in the \Planck\ 353--857\,GHz and DIRBE 240--60 $\mu$m channels; at
25--12\,$\mu$m it accounted for more than 80\% of the signal variance.

A notable short-coming of the analysis of \citet{CG02_05} was its low
angular resolution, determined by the $7^{\circ}$ FWHM FIRAS
beam. Aiming to improve on this issue, \citet{CG02_06} therefore
derived high-resolution maps of the cold (HI-correlated) and hot
(CII-correlated) components through a high-dimensional posterior grid
search. Together with the \GAIA\ extinction template
\citet{edenhofer:2024} and the H$\alpha$ template \citep{wham:2003},
which are natively provided with a resolution appropriate for DIRBE
analysis, the resulting set of four templates provides a useful basis
set for modelling dust throughout the DIRBE frequency range. These
templates are plotted in \cref{fig:templates}.

While analyzing the HFI data with this model, \citet{CG02_06} found
that the SED of all four components could be very well approximated by
MBB spectra throughout the HFI frequency range. However, when
extending into the higher-frequency DIRBE range, non-thermal physics
becomes increasingly important, and a simple parametric SED is no
longer valid. Instead, for the DIRBE frequency range we define a set
of SED bins, each of which is roughly chosen to correspond to the width of a
DIRBE band, as shown in \cref{tab:bands}. Each dust component is then
defined to have a constant amplitude within a given bin, resulting in one
free parameter per bin per component.

Based on this model, the \Cosmoglobe\ DR2 sky model (i.e., the third
term of \cref{eq:model}) takes the following form,
\begin{equation}
\label{eq:skymodel}
\begin{aligned}
  \sum_{c=1}^{n_{\mathrm{comp}}} \M_c(\nu) \a_c  = 
    &[a]_{\mathrm{cold}}(\nu)\t_{\mathrm{cold}} && \textrm{(Cold dust)} \\
    + & [a]_{\mathrm{hot}}(\nu)\t_{\mathrm{hot}} && \textrm{(Hot dust)}\\
    + & [a]_{\mathrm{nearby}}(\nu)\t_{\mathrm{nearby}} && \textrm{(Nearby dust)} \\
    + & [a]_{\mathrm{H\alpha}}(\nu)\t_{\mathrm{H\alpha}} && \textrm{(H$\alpha$ correlated dust)} \\
  + &\left(\frac{\nuzeroff}{\nu}\right)^2
  \frac{g_{\mathrm{ff}}(\nu;\Te) }{g_{\mathrm{ff}}(\nuzeroff;\Te)}
  \vec{t}_{\mathrm{ff}} && \textrm{(Free-free)} \\
  + &U_{\mathrm{mJy}} \sum_{j=1}^{n_{\mathrm{s}}}
  f_{\mathit{Gaia},j} a_{\mathrm{s},j} &\quad&
  \textrm{(Bright stars)} \\
  + &U_{\mathrm{mJy}} f_{\mathit{Gaia},j} \a_{\mathrm{fs},j} &\quad&
  \textrm{(Faint stars)} \\  
  + &m_{\nu} && \textrm{(Monopole)}. 
\end{aligned}
\end{equation}
In this equation, the bracketed amplitudes, $[a](\nu)$, indicate the
binned SED amplitudes discussed above, while $\t_i$ represents the
corresponding template maps as shown in Fig.~\ref{fig:templates}. The
first four terms in this equation form our model of thermal dust,
which is the main focus of this work. The fifth term models the
free-free emission, which contributes only a small contribution at all
relevant frequencies. The sixth and seventh terms represent two point
source contributions (primarily stars), which are described in detail
by \citet{CG02_04}. Finally, the last term represents the zero-level
at each frequency, which ideally should represent the CIB monopole
\citep{CG02_03}.

\begin{table}[t]
    \centering
    \caption{Components enabled for each frequency band. The dust band widths represent the width of each dust band used in this analysis, not the instrumental bandwidths.}
    \begin{tabular}{c|c|c|c|c|c|c}
        \label{tab:bands}
         & Bin width &  &  &  & \\
        Band & (GHz) & Hot & Cold & Nearby & H$\alpha$ \\
        \hline
        DIRBE 240 $\mu\mathrm m$ & 617   &\x &\x &\x &\x \\
        DIRBE 140 $\mu\mathrm m$ & 873   &\x & &\x &\x \\
        DIRBE 100 $\mu\mathrm m$ & 1524  &\x & &\x &\x \\
        DIRBE 60  $\mu\mathrm m$ & 4936  &\x & &\x &\x \\
        DIRBE 25  $\mu\mathrm m$ & 9100  &\x & &\x &\x \\
        DIRBE 12  $\mu\mathrm m$ & 27400 &\x & &\x &\x \\
        DIRBE 4.9 $\mu\mathrm m$ & 24715 &\x & &\x & \\
        DIRBE 3.5 $\mu\mathrm m$ & 39275 &\x & &\x & \\
    \end{tabular} 
\end{table}

\subsection{Bayesian end-to-end analysis}

Taking into account both the general data model in Eq.~\ref{eq:model}
and the sky model in Eq.~\ref{eq:skymodel}, the full set of free
parameters in the \Cosmoglobe\ DR2 analysis is $\omega=\{\G,
\zeta_{\mathrm{zodi}}, \s_{\mathrm{static}}, [a]_{\mathrm{cold}},
     [a]_{\mathrm{hot}},
     [a]_{\mathrm{nearby}},[a]_{\mathrm{H}\alpha},a_{s,i},
     \xi_{\mathrm{c}}, m_\nu\}$,
     and the goal is now to estimate these jointly by drawing samples
     from the posterior distribution \citep{CG02_01} --- that is, we
     aim to map out $P(\omega | \dv)$, the probability distribution of
     the set of parameters $\theta$ given the observed data
     $\dv$. Bayes' theorem allows us to write this on the following
     form,
\begin{equation}
    \label{eq:bayestheorem}
    P(\omega|\dv) = \frac{P(\dv|\omega)P(\omega)}{P(\dv)}.
\end{equation}
In this equation, $P(\dv|\omega)$ is called the likelihood;
$P(\omega)$ is called the prior; and $P(\dv)$ is called the
evidence. While the latter plays an important role in Bayesian model
selection applications, we are in the current work only concerned with
parameter optimization, and the evidence is then just a normalization
constant. 

The number of parameters involved in $\omega$ is large, and the
correlations between the various components are complicated, and this
makes sampling from the posterior function a highly non-trivial
task. The \Cosmoglobe\ framework is based on the \Commander\ software
\citep{eriksen:2004,seljebotn:2019,bp03}, which maps out the posterior
parameter distribution through a process called Gibbs sampling
\citep[e.g.,][]{geman:1984}, a Monte-Carlo method based on
sequentially sampling each parameter (or a subset of parameters) from
their respective marginal distributions with respect to all other
parameters. The theory of Gibbs sampling then guarantees that by
iterating over all conditional distributions, the resulting
multi-variate sample sets will represent a proper sample from the true
joint distribution. For the Cosmoglobe DR2 analysis, the Gibbs chain
looks as follows:
%\begin{equation}
%\begin{alignat}{11}
%\G &\,\leftarrow P(\G&\,\mid &\,\dv,&\, &\,\phantom{\G} &\,\xi_n, &
%\,\beta_{\mathrm{sky}}& \,\a_{\mathrm{sky}}, &\,\zeta_{\mathrm{z}},
%&\,\a_{\mathrm{static}})\\ \nonumber
%\xi_{\mathrm{n}} &\,\leftarrow P(\xi_{\mathrm{n}}&\,\mid &\,\dv,&\, &\,\G, &\,\phantom{\xi_n} &
%\,\beta_{\mathrm{sky}}& \,\a_{\mathrm{sky}}, &\,\zeta_{\mathrm{z}},
%&\,\a_{\mathrm{static}})\\ \nonumber
%\beta_{\mathrm{sky}} &\,\leftarrow P(\beta_{\mathrm{sky}}&\,\mid &\,\dv,&\, &\,\G, &\,\xi_n, &
%\,\phantom{\beta_{\mathrm{sky}}}& \,\a_{\mathrm{sky}}, &\,\zeta_{\mathrm{z}}, &\,\a_{\mathrm{static}})\\ \nonumber
%\a_{\mathrm{sky}} &\,\leftarrow P(\a_{\mathrm{sky}}&\,\mid &\,\dv,&\, &\,\G, &\,\xi_n, &
%\,\beta_{\mathrm{sky}},& \,\phantom{\a_{\mathrm{sky}},}
%&\,\zeta_{\mathrm{z}}, &\,\a_{\mathrm{static}})\\ \nonumber
%\zeta_{\mathrm{z}} &\,\leftarrow P(\zeta_{\mathrm{z}}&\,\mid &\,\dv,&\, &\,\G, &\,\xi_n, &
%\,\beta_{\mathrm{sky}},& \,\a_{\mathrm{sky}},
%&\,\phantom{\zeta_{\mathrm{z}},} &\,\a_{\mathrm{static}})\\ \nonumber
%\a_{\mathrm{static}} &\,\leftarrow P(\a_{\mathrm{static}}&\,\mid &\,\dv,&\, &\,\G, &\,\xi_n, &
%\,\beta_{\mathrm{sky}},& \,\a_{\mathrm{sky}}, &\,\zeta_{\mathrm{z}} &\,\phantom{\a_{\mathrm{static}}})\label{eq:gibbs_static}.
%\end{alignat}
%\end{equation}
\begin{equation}
    \label{eq:gibbschain}
\begin{aligned}
\G &\,\leftarrow P(\G&\,\mid &\,\dv,&\, &\,\phantom{\G} &\,\xi_n, &
\,[a]_i,& \,\a_{\mathrm{s},i}, &\,\zeta_{\mathrm{z}},
&\,\a_{\mathrm{static}})\\
\xi_{\mathrm{n}} &\,\leftarrow P(\xi_{\mathrm{n}}&\,\mid &\,\dv,&\, &\,\G, &\,\phantom{\xi_n} &
\,[a]_i,& \,\a_{\mathrm{s},i}, &\,\zeta_{\mathrm{z}},
&\,\a_{\mathrm{static}})\\
[a]_i &\,\leftarrow P([a]_i&\,\mid &\,\dv,&\, &\,\G, &\,\xi_n, &
\,\phantom{[a]_i}& \,\a_{\mathrm{s},i}, &\,\zeta_{\mathrm{z}}, &\,\a_{\mathrm{static}})\\
\a_{\mathrm{s},i} &\,\leftarrow P(\a_{\mathrm{s},i}&\,\mid &\,\dv,&\, &\,\G, &\,\xi_n, &
\,[a]_i,& \,\phantom{\a_{\mathrm{s},i},}
&\,\zeta_{\mathrm{z}}, &\,\a_{\mathrm{static}})\\
\zeta_{\mathrm{z}} &\,\leftarrow P(\zeta_{\mathrm{z}}&\,\mid &\,\dv,&\, &\,\G, &\,\xi_n, &
\,[a]_i,& \,\a_{\mathrm{s},i},
&\,\phantom{\zeta_{\mathrm{z}},} &\,\a_{\mathrm{static}})\\
\a_{\mathrm{static}} &\,\leftarrow P(\a_{\mathrm{static}}&\,\mid &\,\dv,&\, &\,\G, &\,\xi_n, &
\,[a]_i,& \,\a_{\mathrm{s},i}, &\,\zeta_{\mathrm{z}} &\,\phantom{\a_{\mathrm{static}}})
\end{aligned}
\end{equation}
Here, the symbol $\leftarrow$ indicates the operation of drawing a sample from
the distribution on the right-hand side (see \citet{CG02_01} for a complete
overview of all the symbols in this equation) . After some burn-in period, the
resulting joint parameter sets will correspond to samples drawn from the true
underlying joint posterior.

Since every step of the Gibbs sampling process assumes that all other
parameters are ``given'', we can now treat the fully interconnected
problem (i.e., sampling from the joint posterior of all parameters
involved in our data model) as a modular one -- meaning that we can
perform each ``sub-analysis'' without being concerned with the other
parts of the problem.  Hence, in this paper, we mainly focus on the
four first components of Eq.~(\ref{eq:skymodel}), leaving the
treatment of stars to \citet{CG02_04}; monopoles to \citet{CG02_03};
zodiacal light to \citet{CG02_02}; and condition on free-free emission
as determined by \citet{planck2014-a12}.

\begin{figure}[t]
  \centering
  \includegraphics[width=\columnwidth]{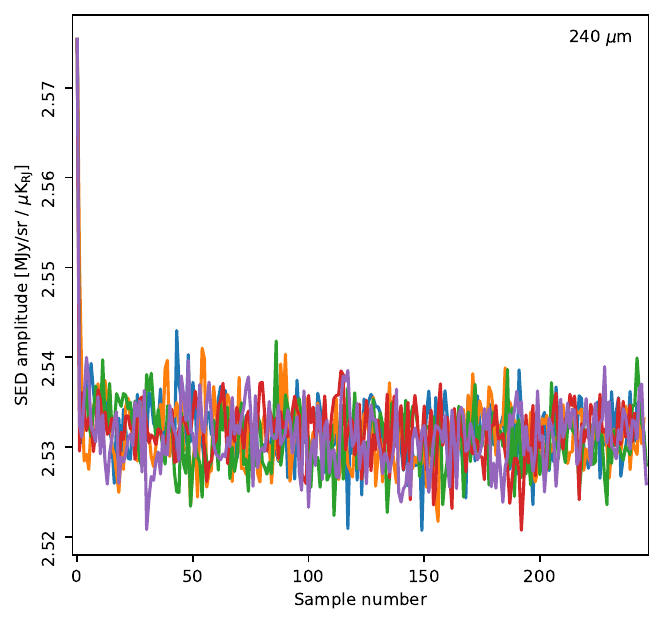}
  \caption{Cold dust amplitude as a function of iteration for the 240 $\mu$m channel where it is included. The five lines correspond to the five independent sampling chains in the analysis. We see robust mixing in all chains.}
  \label{fig:trace_colddust}
\end{figure}

\begin{figure}[t]
  \centering
  \includegraphics[width=\columnwidth]{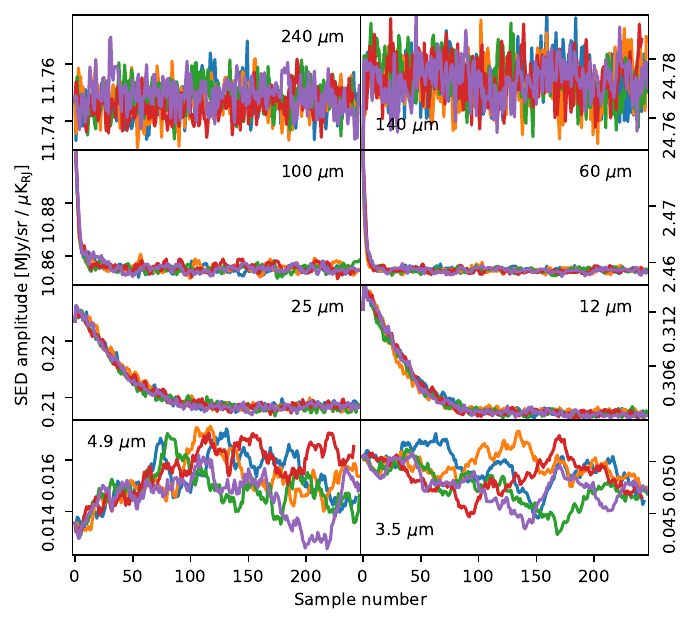}
  \caption{Hot dust amplitudes as a function of iteration for the eight lowest frequency DIRBE channels, with all five sampling chains overplotted. We see that the 3.5 $\mu$m and 4.9 $\mu$m channels exhibit slower mixing than the others, but still manage to explore the full parameter space.}
  \label{fig:trace_hotdust}
\end{figure}
\begin{figure}[t]
  \centering
  \includegraphics[width=\columnwidth]{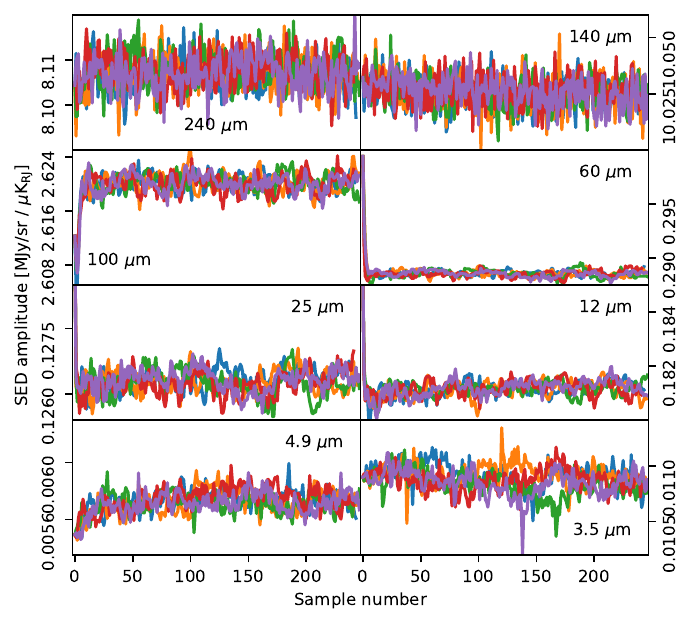}
  \caption{Nearby dust amplitudes as a function of the iteration for the six lowest frequency DIRBE channels, with all five sampling chains overplotted. We see robust mixing in all chains and for all frequency channels.}
  \label{fig:trace_nearbydust}
\end{figure}
\begin{figure}[t]
  \centering
  \includegraphics[width=\columnwidth]{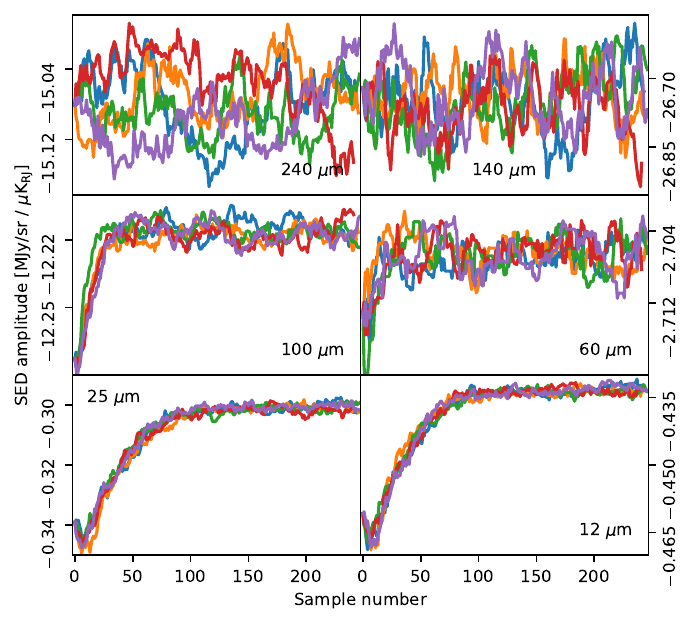}
  \caption{H$\alpha$ dust amplitudes as a function of the iteration for the six
	lowest frequency DIRBE channels, with all five sampling chains
	overplotted. The 240 and 140\,$\mu$m channels exhibit slower mixing than the others, but still manage to explore the full parameter space.}
  \label{fig:trace_hacorr_dust}
\end{figure}

As far as the current paper is concerned, the critical step is
therefore to be able sample from $P([a]_i | \dv, \ldots)$. The actual
likelihood for this distribution is given by first subtracting all other
physical components from the raw data, and then noting that the
residual should ideally be given by white noise. The appropriate
likelihood is therefore the following Gaussian,
\begin{equation}
\mathcal{L}([a]_i) \equiv P(\dv|[a]_i, \ldots) \propto e^{-\frac{1}{2} \sum_{\nu}\left(\frac{\r_\nu -
    [a]_i \t_i}{\sigma_\nu}\right)^2},
\label{eq:likelihood}
\end{equation}
where $\r_{\nu}$ denotes the data-minus-signal residual obtained by
subtracting all other model components except the currently sampled
one, and $\sigma_\nu$ is the instrumental noise rms per TOD sample. We
impose no prior on the amplitudes, and therefore the conditional
distribution is identical to this likelihood.

To sample from this distribution, we implement a standard Metropolis
MCMC accept-reject sampler. That is, at each step we propose an
updated value of $[a]_i$, with a small pre-defined step size; evaluate
Eq.~\ref{eq:likelihood}; and accept the new point with probability $a
=
\textrm{min}(\mathcal{L}_{\mathrm{new}}/\mathcal{L}_{\mathrm{old}},1)$. To
obtain reasonable accept rates, we typically choose a large step size
in preliminary exploratory runs; terminate the analysis when no
further improvements are made; and restart from the previous best-fit
solution with a smaller Metropolis step size.

\section{Data}
\label{sec:data}
There are three main datasets used directly in the \Cosmoglobe\ DR2
analysis, namely low-level data from \COBE-DIRBE, starlight positions
from \WISE\ \citep{wright:2010}, and starlight parameters from
\GAIA\ \citep{gaia:2016,gaia:2018}. These are supplemented by our own
dust templates, which are drawn from the analysis performed in
\citet{CG02_06}, and incorporate data from \Planck\ HFI and
\Gaia. Below we give a succinct description of these data sets and the
preprocessing performed; for a more in-depth description, see
\citet{CG02_01} and \citet{CG02_04}.

\subsection{\COBE-DIRBE}
The Diffuse InfraRed Explorer (DIRBE), whose main goal was the mapping
out of the cosmic infrared background, was part of the
\COBE\ satellite \citep{boggess92, silverberg93}, and measured the sky
in ten frequency bands from $1.25\,\mathrm{\mu m}$ to
$240\,\mathrm{\mu m}$. In this analysis, we have converted the
original DIRBE CIO (Calibrated Individual Observations), whose
pointing information is given in terms of Quadcube pixels with a
resolution of 20\arcs\, into HEALPix \nside=512 pixelation maps,
resulting in an approximate angular resolution of 42\arcm. Following
the nomenclature of \cite{CG02_01}, we refer to the DIRBE CIOs as
``time-ordered data'' (TOD). Performing our analysis on DIRBE TOD
directly allows us to target the zodiacal light \citep{CG02_02},
which, as mentioned above, must be treated as a time-dependent sky
component, in contrast to other astrophysical sources.

%The use of DIRBE data also allows a fuller exploration of the scales
%relevant for interstellar dust modelling; in particular, we are able
%to capture the peak of the MBB's and their exponential fall-off at
%shorter wavelengths. Although these data introduce additional modeling
%complexity, namely PAH emission, they can break key degeneracies
%between $T$ and $\beta$.

\begin{figure}[t]
  \centering
  \includegraphics[width=\columnwidth]{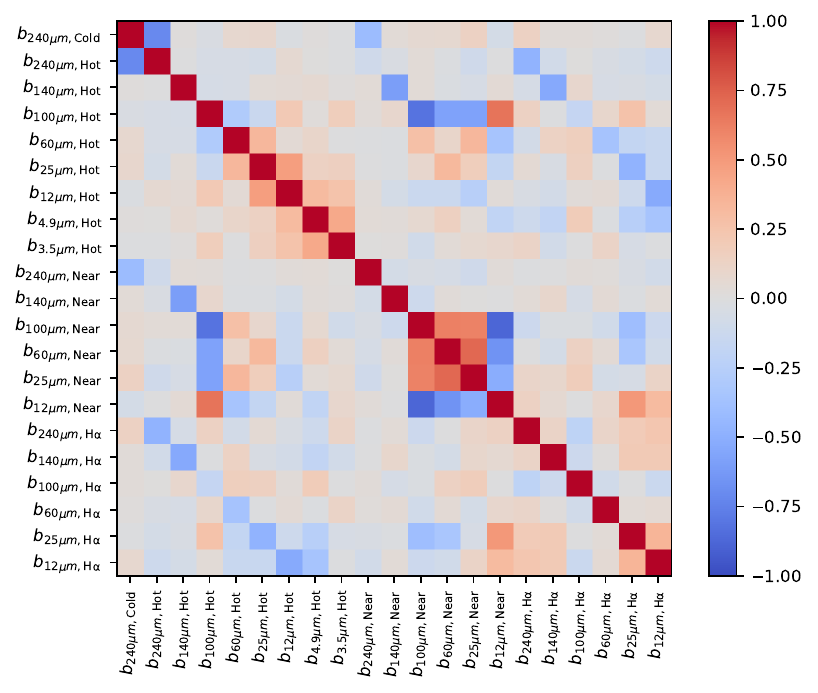}
  \caption{Correlations between the dust amplitudes at each frequency, as
	computed over the full sample set. They are largely
	uncorrelated, but there are some structures within some bands as the
	sampler trades off between components (e.g., 240\,$\mu$m in the upper
	left). There is also some fainter structure in the nearby dust amplitudes across channels.}
  \label{fig:corrmat}
\end{figure}

\begin{table}[t]
\newdimen\tblskip \tblskip=5pt
\caption{Summary of dust template amplitude posterior constraints.}
\label{tab:tempamp}
\vskip -4mm
\footnotesize
\setbox\tablebox=\vbox{
 \newdimen\digitwidth
 \setbox0=\hbox{\rm 0}
 \digitwidth=\wd0
 \catcode`*=\active
 \def*{\kern\digitwidth}
  \newdimen\dpwidth
  \setbox0=\hbox{.}
  \dpwidth=\wd0
  \catcode`!=\active
  \def!{\kern\dpwidth}
  \halign{\hbox to 1.7cm{#\leaderfil}\tabskip 2em&
    \hfil$#$\hfil \tabskip 0.5em&
    \hfil$#$\hfil \tabskip 1em\cr
\omit\sc Component (Thermal dust band) &  \omit\sc Template amplitude [MJy/sr] \hfil\cr
\noalign{\doubleline}
Cold dust (240 $\mu$m) & 131.03 \pm 0.16 \cr
H$\alpha$-correlated dust (240 $\mu$m) & -26.853 \pm 0.065 \cr
H$\alpha$-correlated dust (140 $\mu$m) & -47.613 \pm 0.119 \cr
H$\alpha$-correlated dust (100 $\mu$m) & -21.754 \pm 0.007 \cr
H$\alpha$-correlated dust (60 $\mu$m) & -4.819 \pm 0.003 \cr
H$\alpha$-correlated dust (25 $\mu$m) & -0.536 \pm 0.002 \cr
H$\alpha$-correlated dust (12 $\mu$m) & -0.772 \pm 0.001 \cr
Nearby dust (240 $\mu$m) & 392.383 \pm 0.182 \cr
Nearby dust (140 $\mu$m) & 485.184 \pm 0.336 \cr
Nearby dust (100 $\mu$m) & 126.808 \pm 0.062 \cr
Nearby dust (60 $\mu$m) & 13.964 \pm 0.015 \cr
Nearby dust (25 $\mu$m) & 6.113 \pm 0.014 \cr
Nearby dust (12 $\mu$m) & 8.770 \pm 0.020 \cr
Nearby dust (4.9 $\mu$m) & 0.278 \pm 0.003 \cr
Nearby dust (3.5 $\mu$m) & 0.526 \pm 0.005 \cr
Hot dust (240 $\mu$m) & 393.585 \pm 0.180 \cr
Hot dust (140 $\mu$m) & 829.926 \pm 0.242 \cr
Hot dust (100 $\mu$m) & 363.674 \pm 0.045 \cr
Hot dust (60 $\mu$m) & 82.369 \pm 0.012 \cr
Hot dust (25 $\mu$m) & 6.981 \pm 0.017 \cr
Hot dust (12 $\mu$m) & 10.078 \pm 0.009 \cr
Hot dust (4.9 $\mu$m) & 0.513\pm 0.033 \cr
Hot dust (3.5 $\mu$m) & 1.609 \pm 0.063 \cr }}
\endPlancktablewide
\end{table}

\begin{figure}[t]
  \centering
  \includegraphics[width=\columnwidth]{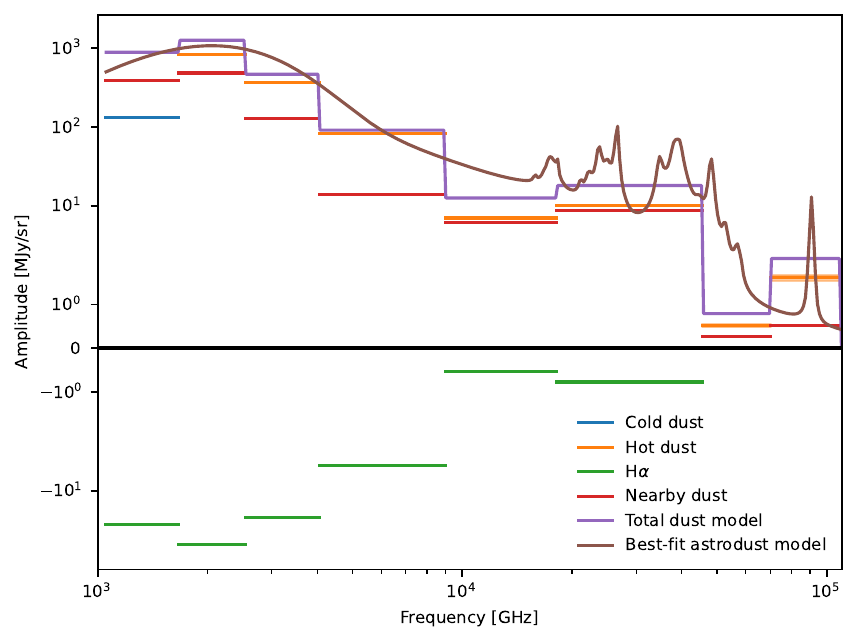}
  \caption{The total dust SED as a function of frequency, as well as the four
      dust components. The best-fit astrodust model fit to the total SED for
  the frequencies shown is also plotted in brown.}
  \label{fig:total_sed}
\end{figure}

\subsection{\WISE\ and \GAIA}
At the higher DIRBE frequency bands (25--1.25 $\mu$m), starlight
emission becomes a significant source of emission, both as point
sources and as a diffuse component. Using the AllWise point source
catalog \citep{CatWISE}, we crossmatched the brighest stars in the
\textit W1 band against \Gaia\ DR2, and used the estimated physical
parameters from that catalogue to model the bright stars. In addition,
we created a general faint source template, similar to that used in
\citet{hauser1998}, based on the stars that were not part of the
bright star component (see \citet{CG02_04} for more
details). Together, these two components comprise the sixth and
seventh component in \cref{eq:skymodel}.

\subsection{Data selection}
As noted in \citet{CG02_05}, not all thermal dust components are
expected to contribute to the DIRBE bands we are considering in this
paper (i.e., 240--12$\mu$m).  Following the same logic as in that
paper, we restrict the dust components to be active in the various
DIRBE bands as indicated in \cref{tab:bands}. In particular, the cold
dust component peaks at 1\,THz, and falls off quickly at higher
frequencies, making it quicly subdominant to all other components.  In
contrast, the hot dust component is visible until the
3.5\,$\mathrm{\mu m}$ channel, and must be included to obtain a
satisfactory fit. The strength of the nearby component is
intermediate, while the impact of the H$\alpha$-correlated components
is restricted to a relatively small fraction of the sky. 

\begin{figure*}[t]
  \centering
  \includegraphics[width=0.235\linewidth]{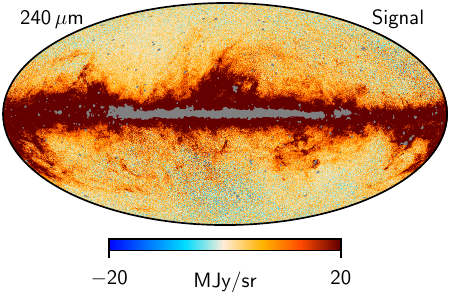}
  \includegraphics[width=0.235\linewidth]{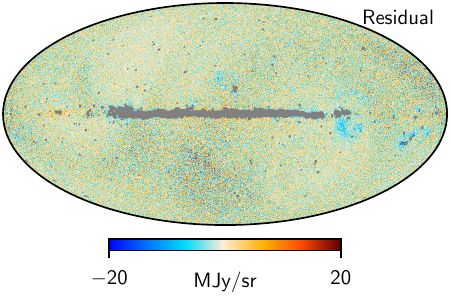}\hspace*{5mm}
  \includegraphics[width=0.235\linewidth]{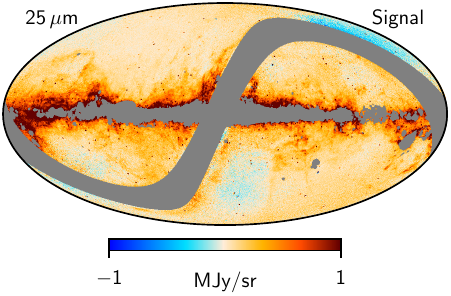}
  \includegraphics[width=0.235\linewidth]{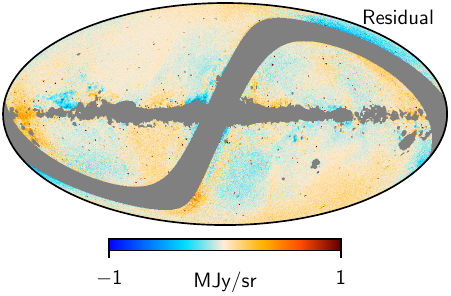}\\
  \includegraphics[width=0.235\linewidth]{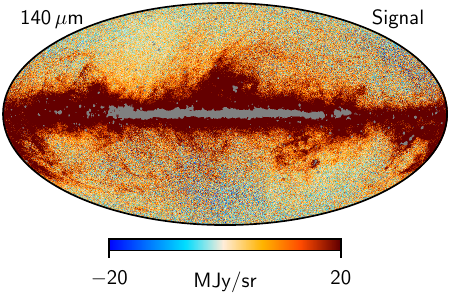}
  \includegraphics[width=0.235\linewidth]{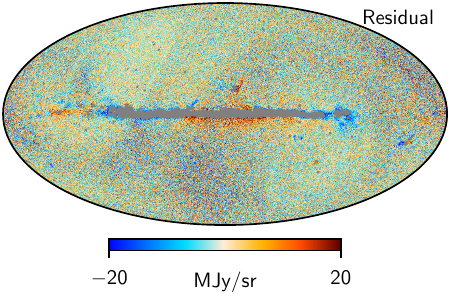}\hspace*{5mm}
  \includegraphics[width=0.235\linewidth]{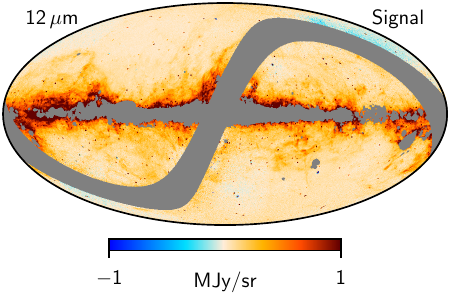}
  \includegraphics[width=0.235\linewidth]{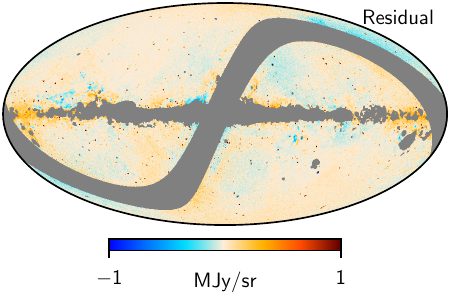}\\
  \includegraphics[width=0.235\linewidth]{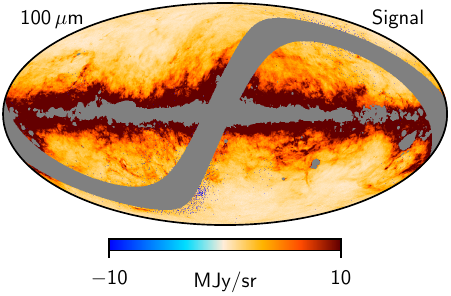}
  \includegraphics[width=0.235\linewidth]{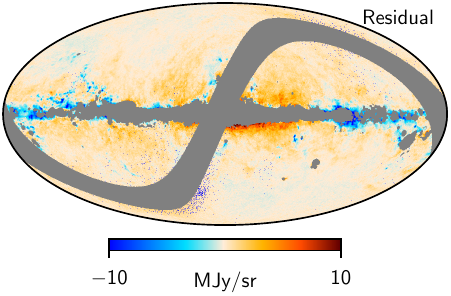}\hspace*{5mm}
  \includegraphics[width=0.235\linewidth]{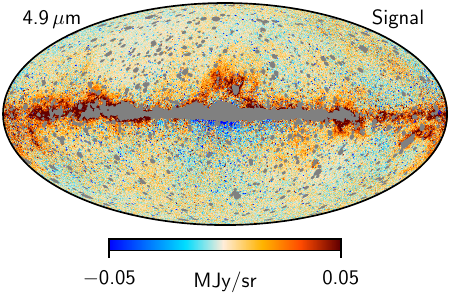}
  \includegraphics[width=0.235\linewidth]{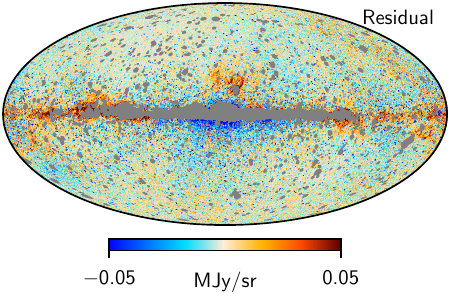}\\
  \includegraphics[width=0.235\linewidth]{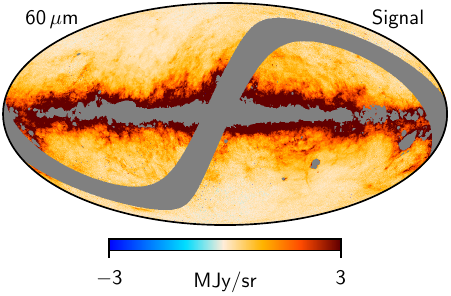}
  \includegraphics[width=0.235\linewidth]{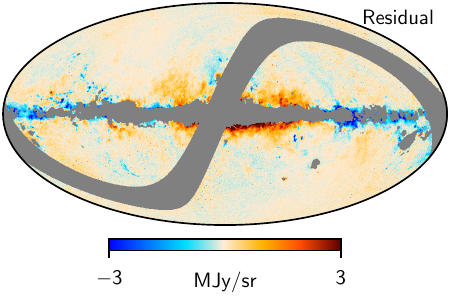}\hspace*{5mm}
  \includegraphics[width=0.235\linewidth]{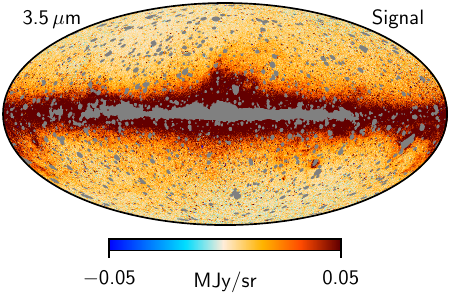}
  \includegraphics[width=0.235\linewidth]{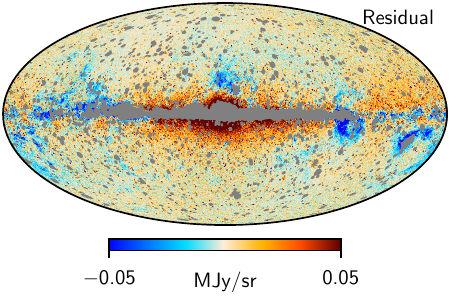}
  \caption{Comparison between thermal dust frequency maps (i.e., stationary sky signal minus starlight and free-free; \emph{first and third columns}) and residual maps for each frequency channel (\emph{second and fourth column}). Gray pixels indicate the analysis masks used for each frequency channel.}
  \label{fig:dustmaps}
\end{figure*}

\begin{figure}[t]
  \centering
  \includegraphics[width=\columnwidth]{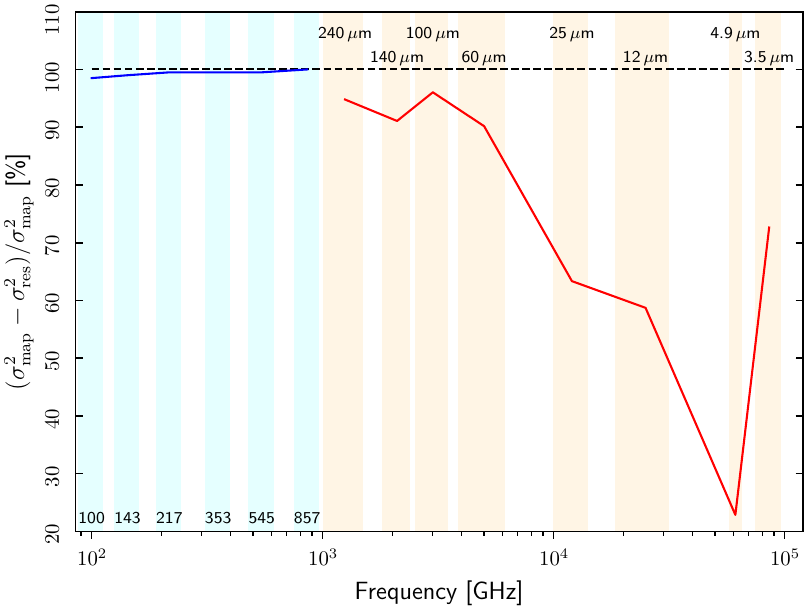}
  \caption{Dust model efficiency as a function of frequency, as defined in terms of variance reduction. The red line shows results for DIRBE, as evaluated from the maps shown in \cref{fig:dustmaps}, while the blue line shows results for \Planck\ HFI, as presented by \citet{CG02_06}. Vertical bands indicate the position and bandwidth of each DIRBE (orange) and HFI (cyan) frequency channel.}
  \label{fig:efficiency}
\end{figure}

%\begin{figure}[t]
%  \centering
%  \includegraphics[width=\linewidth]{figures/paperVII_07_todres_zoom_v1.pdf}\\
%  \includegraphics[width=\linewidth]{figures/paperVII_08_todres_zoom_v1.pdf}
%	\caption{Full-sky data-minus-model residual maps for the 60 (\emph{top}) and 100\,$\mu$m (\emph{bottom}) DIRBE channels. }
%  \label{fig:res0708}
%
%\end{figure}

\subsection{Masks}
\label{sec:masks}
As discussed by \citet{CG02_01}, analysis masks are generated based on
dust residual amplitudes and zodiacal dust residuals. For the dust
components, regions that are poorly modeled are excluded depending on
frequency and residual amplitude. This is determined using a
combination of smoothed residual maps and $\chi^2$ maps per frequency
channel. This ensures that only data that are poorly modeled are
masked, while retaining well-modeled high signal-to-noise pixels.  For
the bands where zodiacal dust is brightest (12--100\,$\mathrm{\mu
  m}$), we apply an additional mask along the Ecliptic plane. This
mask was determined solely by the brightest zodiacal dust channel,
$25\,\mathrm{\mu m}$, and is largely due to asteroidal band
uncertainty.

\begin{figure*}[t]
  \centering
  \includegraphics[width=0.95\textwidth]{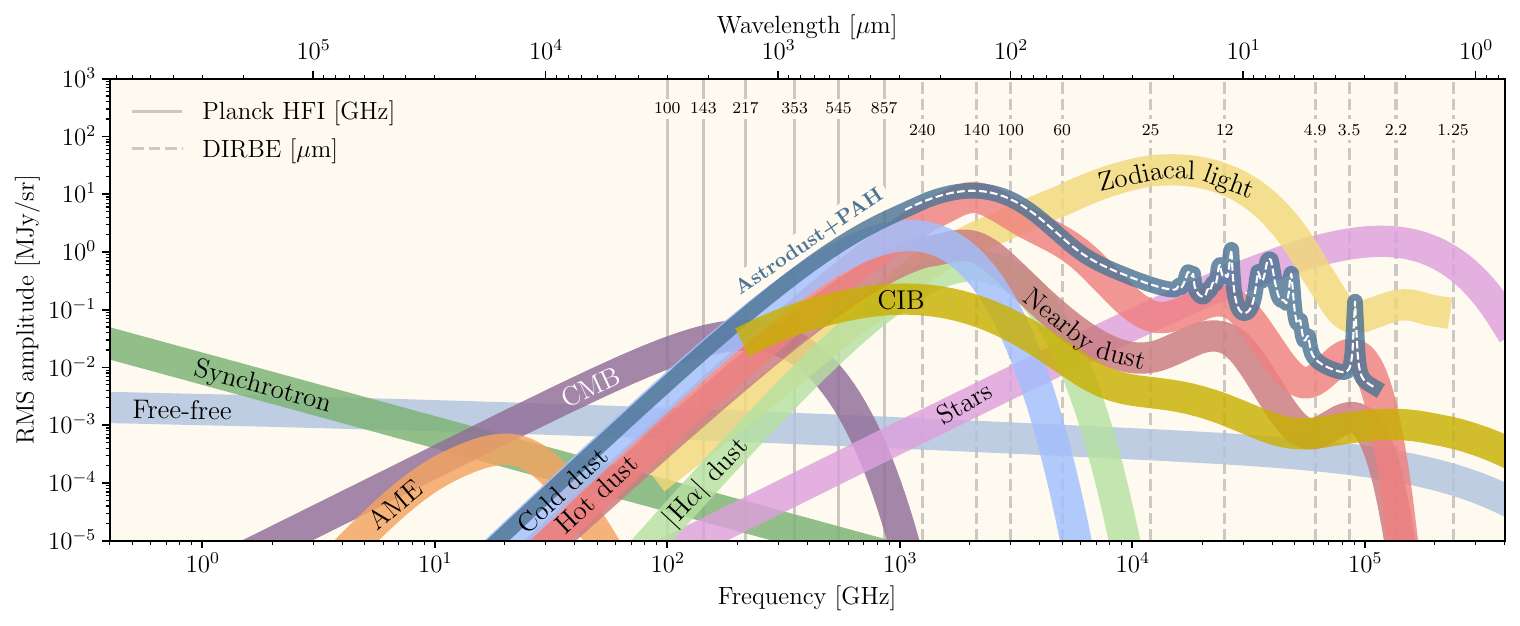}
	\caption{Overview of the large-scale microwave and infrared sky from 1 GHz to 1$\mu$m, based on the \Cosmoglobe\ DR1 and DR2 data. The four component dust model is shown, as well as the best fit astrodust model. The \Planck\ HFI and DIRBE central frequencies are indicated with  vertical lines.}
  \label{fig:SED_overview}
\end{figure*}

%\FloatBarrier
\section{Results}

The \Cosmoglobe\ DR2 production analysis resulted in five Gibbs
chains, each of which contains approximately 250 samples. Out of
these, the first 100 samples are considered part of the burn-in
period, after which all chains have converged. In the following, we
include trace plots with the burnin period included to visualize the
convergence properties of the chains.

\subsection{Markov chains, correlations and convergence}
In
\cref{fig:trace_colddust,fig:trace_hotdust,fig:trace_hacorr_dust,fig:trace_nearbydust},
we plot the Gibbs chain traceplots for all the five Gibbs chains used
in the analysis, for all four dust components. After a period of
reaching equilibrium --- which varies from almost instantly (for the
cold dust amplitude, for example) to around 100~samples at the most
(particularly evident for the hot dust and H$\alpha$-correlated dust
25 and 12\,$\mu $m bins) --- the chains generally do not exhibit any
long correlations and seem to mix fairly well. The only exceptions are
the 240 and 140\,$\mu$m H$\alpha$-correlated dust bins, where we see
more long-term trends that indicate a slower traversal through
parameter space.

In \cref{fig:corrmat}, we show the correlations between the various
SED bins in the run, calculated over all five chains after discarding
burn-in. The strongest correlations are internally between the nearby
dust bins. Most of the bins are correlated with each other, but they
are anticorrelated with the 12\,$\mu$m amplitude. The correlation of
the nearby dust component's amplitude between different bins is due to
indirect interactions of the parameters in the model, as the
morphology of the nearby dust is fixed, and the amplitudes are sampled
separately. 

Those same bins, and in particular the 100\,$\mu$m bin, are also
anticorrelated with the 100\,$\mu$m hot dust amplitude.  This is not
unexpected, as there is morphological overlap between the hot dust and
nearby dust components, as displayed in
\cref{fig:templates}. Therefore, a negative correlation would be
expected to model the same total amplitude in the DIRBE map
itself. The same phenomenon can be seen between hot and cold dust at
240\,$\mu$m.

%\begin{figure}
%  \centering
%  \includegraphics[width=\columnwidth]{figures/traceplots.pdf}
%  \caption{Traceplots.}
%  \label{fig:trace}
%\end{figure}

\subsection{Multi-component thermal dust SED posteriors}

In \cref{tab:tempamp}, we summarize the thermal dust SED posteriors
resulting from the above analysis as the mean value and variance of
the chain samples (including all five chains of the analysis,
discarding burn-in).  Similarly, we plot the posterior mean values per
bin in \cref{fig:total_sed}, where the thickness of the line indicates
the standard deviation of that amplitude.  In the same plot, we also
show the posterior total SED, wherein all four components are summed
up, as well as the Astrodust+PAH model that is the best fit to the
mean total dust SED.\footnote{Specifically, we fit an overall
amplitude $A$, as well as the $\log_{10}{U}$ parameter used in
\citet{Hensley2023}, with the total model to fit being
$A\cdot\mathrm{EM}(\log_{10}{U})$, where $EM$ is their tabulated
function returning the emission as a function of $\log_{10}{U}$.}

\begin{figure*}[t]
         \centering
         \includegraphics[width=0.16\linewidth]{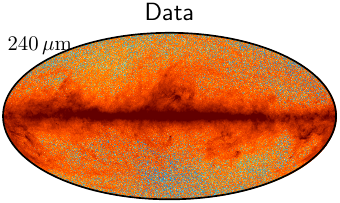}
         \includegraphics[width=0.16\linewidth]{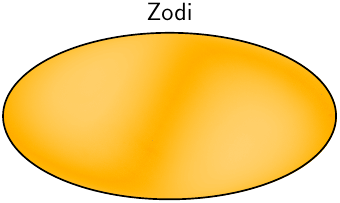}
         \includegraphics[width=0.16\linewidth]{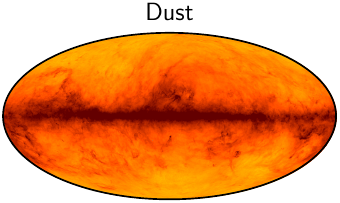}
         \includegraphics[width=0.16\linewidth]{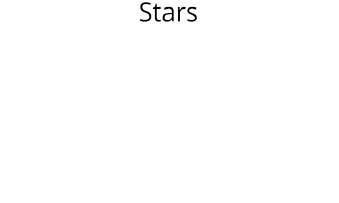}
         \includegraphics[width=0.16\linewidth]{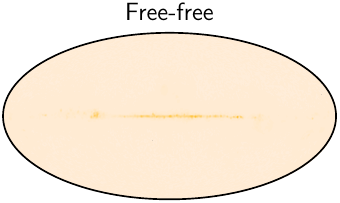}
         \includegraphics[width=0.16\linewidth]{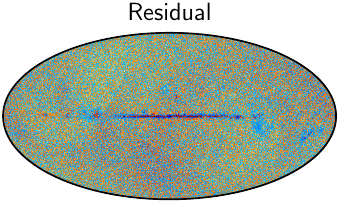}\\
         \includegraphics[width=0.16\linewidth]{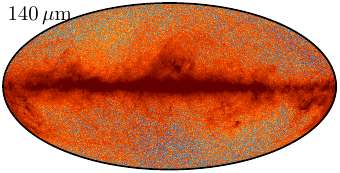}
         \includegraphics[width=0.16\linewidth]{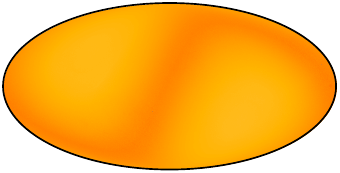}
         \includegraphics[width=0.16\linewidth]{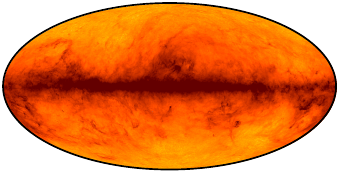}
         \includegraphics[width=0.16\linewidth]{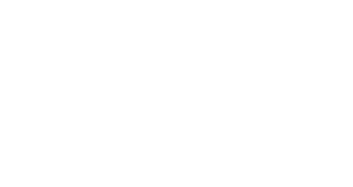}
         \includegraphics[width=0.16\linewidth]{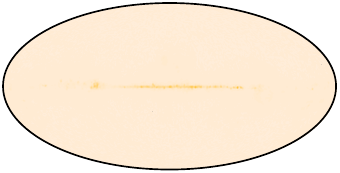}
         \includegraphics[width=0.16\linewidth]{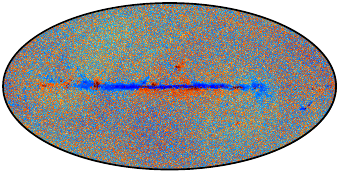}\\
         \includegraphics[width=0.16\linewidth]{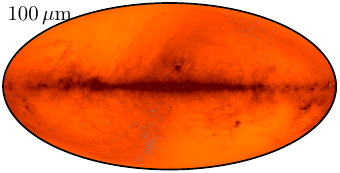}
         \includegraphics[width=0.16\linewidth]{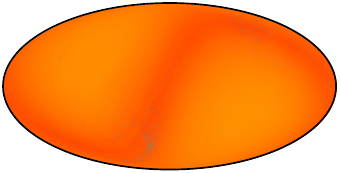}
         \includegraphics[width=0.16\linewidth]{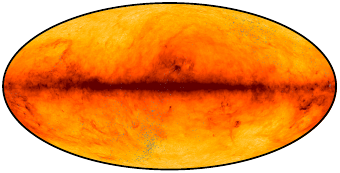}
         \includegraphics[width=0.16\linewidth]{figures/compfreq_white_nobar.pdf}
         \includegraphics[width=0.16\linewidth]{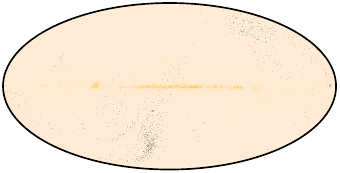}
         \includegraphics[width=0.16\linewidth]{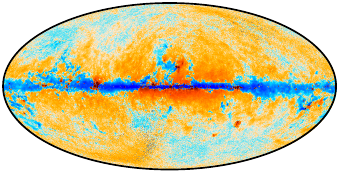}\\
         \includegraphics[width=0.16\linewidth]{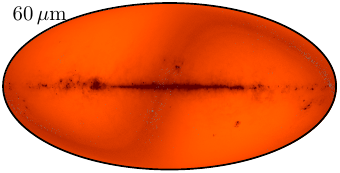}
         \includegraphics[width=0.16\linewidth]{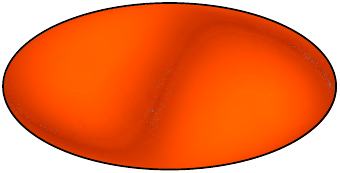}
         \includegraphics[width=0.16\linewidth]{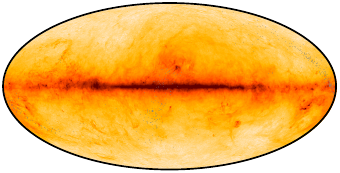}
         \includegraphics[width=0.16\linewidth]{figures/compfreq_white_nobar.pdf}
         \includegraphics[width=0.16\linewidth]{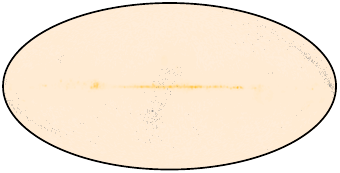}
         \includegraphics[width=0.16\linewidth]{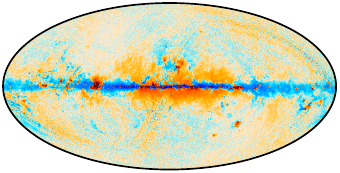}\\
         \includegraphics[width=0.16\linewidth]{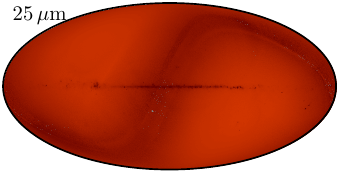}
         \includegraphics[width=0.16\linewidth]{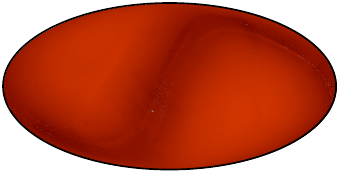}
         \includegraphics[width=0.16\linewidth]{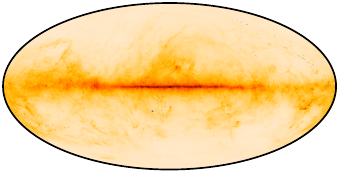}
         \includegraphics[width=0.16\linewidth]{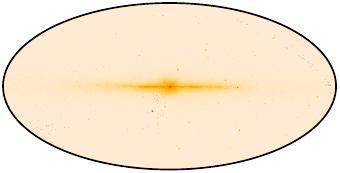}
         \includegraphics[width=0.16\linewidth]{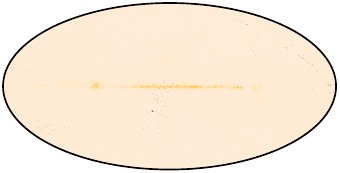}
         \includegraphics[width=0.16\linewidth]{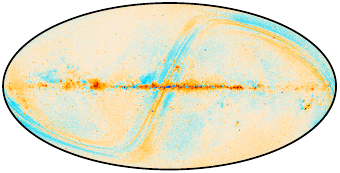}\\  
         \includegraphics[width=0.16\linewidth]{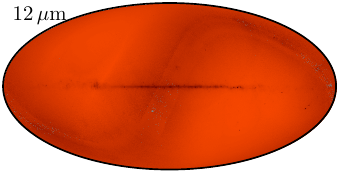}
         \includegraphics[width=0.16\linewidth]{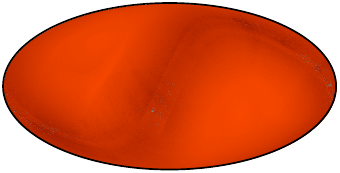}
         \includegraphics[width=0.16\linewidth]{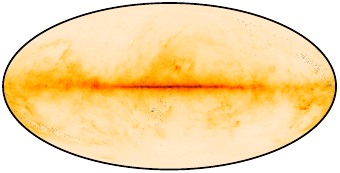}
         \includegraphics[width=0.16\linewidth]{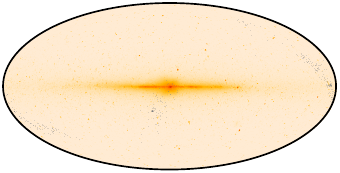}
         \includegraphics[width=0.16\linewidth]{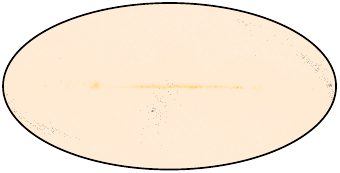}
         \includegraphics[width=0.16\linewidth]{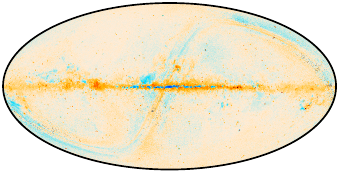}\\
         \includegraphics[width=0.16\linewidth]{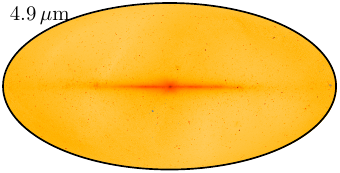}
         \includegraphics[width=0.16\linewidth]{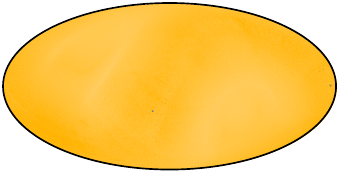}
         \includegraphics[width=0.16\linewidth]{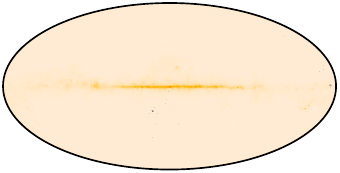}
         \includegraphics[width=0.16\linewidth]{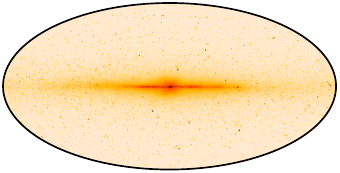}
         \includegraphics[width=0.16\linewidth]{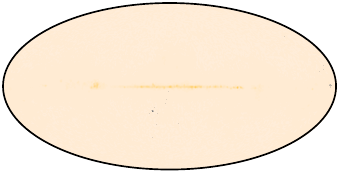}
         \includegraphics[width=0.16\linewidth]{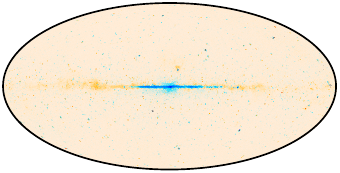}\\  
         \includegraphics[width=0.16\linewidth]{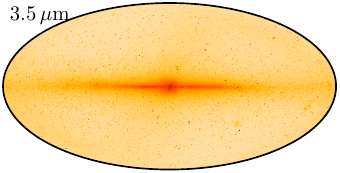}
         \includegraphics[width=0.16\linewidth]{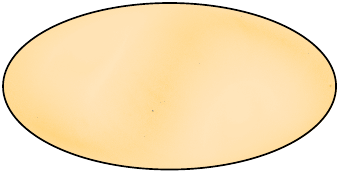}
         \includegraphics[width=0.16\linewidth]{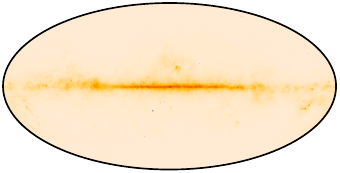}
         \includegraphics[width=0.16\linewidth]{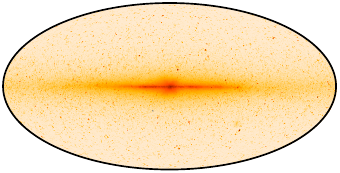}
         \includegraphics[width=0.16\linewidth]{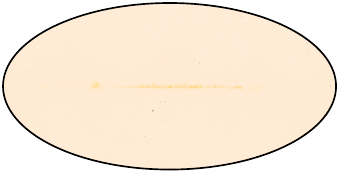}
         \includegraphics[width=0.16\linewidth]{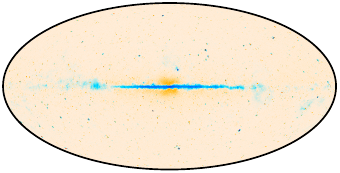}\\  
         \includegraphics[width=0.16\linewidth]{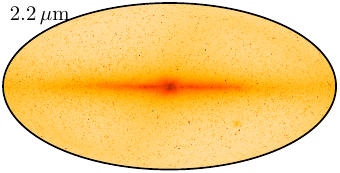}
         \includegraphics[width=0.16\linewidth]{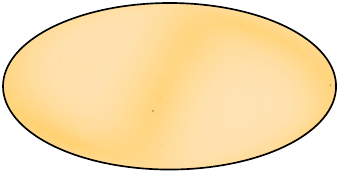}
         \includegraphics[width=0.16\linewidth]{figures/compfreq_white_nobar.pdf}
         \includegraphics[width=0.16\linewidth]{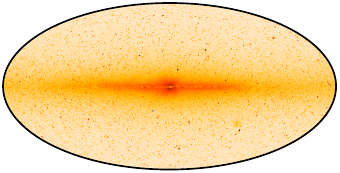}
         \includegraphics[width=0.16\linewidth]{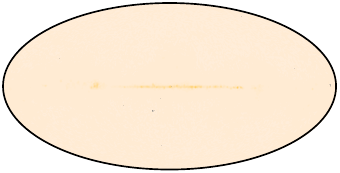}
         \includegraphics[width=0.16\linewidth]{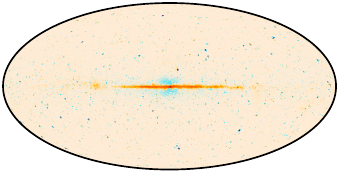}\\
         \includegraphics[width=0.16\linewidth]{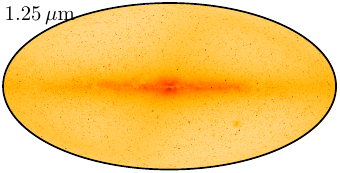}
         \includegraphics[width=0.16\linewidth]{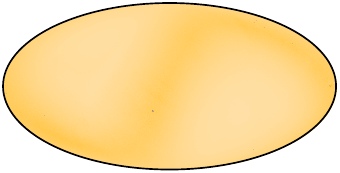}
         \includegraphics[width=0.16\linewidth]{figures/compfreq_white_nobar.pdf}
         \includegraphics[width=0.16\linewidth]{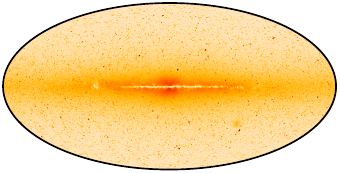}
         \includegraphics[width=0.16\linewidth]{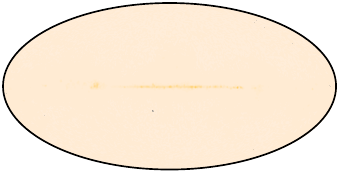}
         \includegraphics[width=0.16\linewidth]{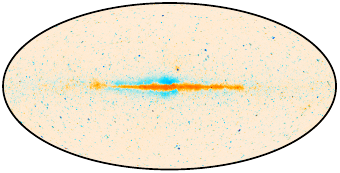}\\
         \includegraphics[width=0.50\linewidth]{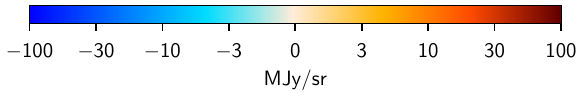}
         \caption{Comparison between the re-analyzed DIRBE data and the various fitted components for one single Gibbs sample. Columns show, from left to right, 1) the time-ordered DIRBE data co-added into pixelized maps; 2) zodiacal light emission; 3) thermal dust emission; 4) star emission; 5) free-free emission; and 6) data-minus-model residual emission. Rows show individual frequency channels. Missing entries corresponds to components that are forced to zero in the model. Note that all panels are plotted with the same color scale in units of MJy/sr, and can be directly compared.}
         \label{fig:comp_vs_freq}
\end{figure*}

Both the hot dust and nearby dust components follow a typical thermal
dust modified blackbody curve up to about 3000\,GHz, mirroring the
results found in \citet{CG02_05}. At higher frequencies, they exhibit
the characteristic rise in the SED, where models like Astrodust+PAH
predicts emission lines from the nanoscale particles. This figure also
shows that fitting the total (i.e., the sum of all the components)
dust model to the Astrodust+PAH model yields a curve that fits very
well with the posterior mean SED values.

We cannot draw any such conclusions about the the cold dust component,
as it is only active for the lowest frequency SED bin. However, the
fact that the amplitude of the cold dust component is about an order
of magnitude smaller than for the hot and nearby dust components at
1\,THz is consistent with the results of \citet{CG02_05}. Future work,
in conjunction with higher sensitivity datasets, such as \IRAS\ and
AKARI, will be useful in further determing the SED of this component.

The H$\alpha$ component, as in the previous analyses, exhibits a
negative SED, and is thus representative of dust extinction rather
than emission. However, it exhibits the same type blackbody spectrum
with a hint of a rise towards the PAH emission region. Again, this is
consistent with the results presented by \citet{CG02_05}.  

%\begin{figure}
%  \centering
%  \includegraphics[width=\columnwidth]{figures/cold_dust_sed.pdf}
%  \caption{Cold dust SED.}
%  \label{fig:cold_dust_sed}
%\end{figure}
%\begin{figure}
%  \centering
%  \includegraphics[width=\columnwidth]{figures/hot_dust_sed.pdf}
%  \caption{Hot dust SED.}
%  \label{fig:hot_dust_sed}
%\end{figure}
%\begin{figure}
%  \centering
%  \includegraphics[width=\columnwidth]{figures/nearby_dust_sed.pdf}
%  \caption{Nearby dust SED.}
%  \label{fig:nearby_dust_sed}
%\end{figure}
%
%\begin{figure}
%  \centering
%  \includegraphics[width=\columnwidth]{figures/nearby_dust_sed.pdf}
%  \caption{H$\alpha$ SED.}
%  \label{fig:wham_sed}
%\end{figure}

\subsection{Model efficiency and residuals}

In \cref{fig:dustmaps}, we show the dust frequency maps --- that is,
the frequency maps with starlight and free-free signal subtracted ---
as well as the residual maps after subtracting the four-component dust
model used in this analysis. The grey pixels indicate the masks used
at each frequency, as described in \cref{sec:masks}. In all bands, we
see a weak pattern of over-and-undersubtractions in the Galactic
plane, suggesting that there is room for further sophistication of the
template model.  For example, the Galactic plane is precisely where we
would expect true spatial variations within a single dust component,
which we have not allowed for in this work. To quantify this, we
estimate the reduction of the RMS of the residuals with respect to the
original map in \cref{fig:efficiency}.

For the 240--60 $\,\mathrm{\mu m}$ channels we find that
$\gtrsim90\,\%$ of the variance is explained by our four-component
model. The first two channels are the cleanest, simply because these
are closest in frequency to the \Planck\ HFI data from which the cold
and hot dust templates are derived, and non-thermal effects have not
yet set in. Between 140 and 60\,$\mathrm{\mu m}$, the model is
clearly struggling to account for emission from the Galactic center,
with similar excesses and oversubtractions, characteristic of
temperature variations.
%This indicates a consistent mismodeled
%component affecting the fit throughout this region. One possible
%contaminant not explicitly modeled in this work is the
%$158\,\mathrm{\mu m}$ \ion{C}{ii} cooling line as measured in, e.g.,
%\citet{1999ApJ...526..207F}. This emission, brightest in the Galactic
%plane, could lead to degeneracies with continuum emission sources,
%biasing results. A full joint analysis with FIRAS data will be
%necessary to properly account for this emission.

For the 25 and 12 $\mathrm{\mu m}$ channels only $\sim$60\,\% of the
signal variance is explaineed by the dust model. This can partially be
explained by incomplete zodiacal dust subtraction, as evidenced by
negative regions in both the signal and residual maps. Much of the
residual structure is uncorrelated with the Galactic plane, indicating
that dust modeling is not the primary cause of the excess residuals.

At 4.9\,$\mu$m, we are limited by the variation of the starlight SEDs,
detailed knowledge of the starlight SEDs, incomplete extinction
corrections, and the resolution of DIRBE \citep{CG02_04}. Because this
work is focused on thermal dust, we leave these residuals for future
work. In particular, the SPHEREx \citep{dore:2014} satellite will be able to resolve and
directly measure the SED of most of the infreared-bright point sources
close to the Galactic plane.

Finally, at $3.5\,\mathrm{\mu m}$ we see that more than 70\,\% of the signal variance is captured by the model. This is despite the fact that starlight is even stronger at 3.5 than at 4.9\,$\mu$m. The explanation is simply the presence of a bright PAH line at 3.3\,$\mu$m. The structure of this may be seen by eye by comparing the 3.5 and 4.9\,$\mu$m signal maps in \cref{fig:dustmaps}; while the 4.9\,$\mu$m map is clearly contamined by over-subtracted starlight emission, the 3.5\,$\mu$m map exhibits a bright Galactic signal with a physically reasonable morphology. That extra power is due to PAH emission, and from the residuals we see that this is also very well modelled by the hot dust template. This observations strongly suggests that correlating recent SPHEREx full-sky observations of the 3.3\,$\mu$m line with \Planck\ HFI and DIRBE far-infrared maps will be fruitful.

%\FloatBarrier
\section{\Cosmoglobe\ DR2 sky model}

Until now, we have focused on the infrared regime of sky modeling,
largely decoupled from the microwave region where \Cosmoglobe\ was
primarily developed.  In this section, we unite the \Planck\ HFI-based
microwave and DIRBE-based infrared regimes to present an integrated
sky model that will serve as a foundation of future global dust
modelling work.

%\subsection{Summary of global sky model}

In \cref{fig:SED_overview} we show the best-fit \Cosmoglobe\ sky model,
derived from the first and second data releases,
including the four-component dust model presented in this paper. Although the
amplitudes chosen are relatively arbitrary (every line of sight in the sky will
potentially have different scalings), the spectral shape of the lines are
directly taken from the Gibbs chains as presented in \citet{watts2023_dr1} and
\citet{CG02_01}. This plot represents the current state of the \Cosmoglobe\
effort to create a common sky model derived with the same data pipeline.

%Thus, we can robustly conclude that both for the \Planck\ HFI and
%DIRBE regimes, dust is ``naturally'' decomposable into a relatively low
%number of constituent components, and that those components tend to correlate
%with traces of various kinds of activity in the Milky Way.

Finally, in \cref{fig:comp_vs_freq}, we show the amplitude maps of all the
components involved in the \Cosmoglobe\ DR2 analysis, including zodiacal light,
dust, starlight, and free-free emission. These maps represent the \Cosmoglobe\
sky model as applied to the DIRBE bands, and in the final column we see the
residual maps at each band. At the two lowest bands, where the thermal dust is
the dominant sky component (see \cref{fig:SED_overview}), the model is
performing quite well (with the same caveats as for the pure dust residual
maps), especially given that we essentially treat the thermal dust as a model
with $\sim$20 free parameters, instead of fitting two spectral parameters per
pixel on the sky.

%\FloatBarrier
\section{Conclusions}
In this paper, which is part of the second \Cosmoglobe\ data release,
we have fitted a four-component dust model to the eight lowest DIRBE
frequency bands. Recent analyses \citep{CG02_05, CG02_06} demonstrated
the feasibility and efficiency of decomposing thermal dust into a
relatively low number of distinct populations with global spectral
parameters.

The model uses a nearby dust template and H$\alpha$ dust template
presented in \citet{CG02_05}, labeled ``nearby dust'', and
``H$\alpha$-correlated dust'', as well as two template components
derived in \citet{CG02_06}, labeled ``Hot dust'', ``Cold dust''
respectively. The SED of each component was defined as a constant bin
over each DIRBE band. We found that the resulting total SED is a good
match to the Astrodust+PAH model, and that the nanoparticle emission
at high frequencies is detected in these components.  As in previous
analyses, the H$\alpha$-correlated component absorbs rather than emits
radiation, and thus has a negative SED.

Finally, we present the current state of the \Cosmoglobe\ sky model,
ranging from a few GHz to $\sim$$10^3$\,THz, and containing diverse
components, such as the CMB, synchrotron radiation, zodiacal light,
starlight, the various thermal dust components, and the CIB. Although
the current results are already very encouraging, this work represents
only a first step towards a global dust model for the radio,
microwave, and infrared regimes. Perhaps the most immediate
improvements will come from incorporating complementary
high-resolution and -sensitivity infrared experiments, for instance
AKARI \citep{murakami:2007}. This will allow the construction of dust
templates that can also be used for analysis of high-resolution CMB
experiments such as Simons Observatory \citep{SO2019}.

Likewise, integrating WISE and SPHEREx data will be a game-changer in
our ability to separate starlight from Galactic emission between 1 and
10\,$\mu$m, as well as unique identification of individual PAH
lines. Given the strong association we have found between the
3.5\,$\mu$m channel and the hot dust component already in this paper,
joint analysis between \Planck\ and SPHEREx stands out as a
potential goldmine for broad-spectrum dust modelling. 
and dust 

These results also raise intruiging questions regarding dust modelling
for CMB $B$-mode experiments that search for the imprint of primordial
gravitational waves. The fact that a simple four-component dust model
works this will for intensity measurements strongly suggests that a
similar approach could work for polarization. In that case, it is
worth considering the prospects of establishing corresponding
polarization templates as the four employed here. First, cold dust
dominates below 1000\,GHz, and most high-precision CMB experiments
will therefore naturally measure this component within their default
frequency range. In order to map hot dust, however, it is important to
have access to higher frequencies, and as shown by \citep{CG02_06}
using HFI data, a maximum frequency of 857\,GHz works very
well. Assessing how much the separation between hot and cold dust
degrades if one only had access to at most, say, 353 or 545\,GHz data
would be very interesting. Moving on to the nearby dust component,
full-sky polarized starlight measurements is key. Such data are
already in the process of being generated by the PASIPHAE experiment
\citep{tassis:2018}, and these could easily become a cornerstone in
the analysis of all future $B$-mode experiments. Finally, the most
challenging template to establish in polarization would be the
H$\alpha$ tracer, simply because H$\alpha$ is expected to be extremely
weakly polarized. At the same time, this does not automatically imply
that impact of the H$\alpha$ template would be negligible in
polarization, simply because it appears in absorption, and the
physical mechanism that causes that effect is just as likely to
absorbe any photon, regardless of their polarization direction. On the
other hand, it is also important to note that the H$\alpha$ template
has a limited effective footprint on the sky, and are mostly
associated with structures at low Galactic latitudes. The affected
regions may therefore most likely be efficiently masked in a future
high-precision $B$-mode analysis.

Overall, we conclude that the current analysis opens up new and
exciting research avenues for global dust modelling from radio to
infrared frequencies. Many of these are already under active
exploration, and will be reported through future \Cosmoglobe\ data
releases.

\begin{acknowledgements}
  We thank Richard Arendt, Tony Banday, Johannes Eskilt, Dale Fixsen,
  Ken Ganga, Paul Goldsmith, Shuji Matsuura, Sven Wedemeyer, Janet
  Weiland and Edward Wright for useful suggestions and guidance.  The
  current work has received funding from the European Union’s Horizon
  2020 research and innovation programme under grant agreement numbers
  819478 (ERC; \textsc{Cosmoglobe}), 772253 (ERC;
  \textsc{bits2cosmology}), 101165647 (ERC, \textsc{Origins}),
  101141621 (ERC, \textsc{Commander}), and 101007633 (MSCA;
  \textsc{CMBInflate}).  This article reflects the views of the
  authors only. The funding body is not responsible for any use that
  may be made of the information contained therein. This research is
  also funded by the Research Council of Norway under grant agreement
  numbers 344934 (YRT; \textsc{CosmoglobeHD}) and 351037 (FRIPRO;
  \textsc{LiteBIRD-Norway}). Some of the results in this paper have been
  derived using healpy \citep{Zonca2019} and the HEALPix
  \citep{healpix} packages.  We acknowledge the use of the Legacy
  Archive for Microwave Background Data Analysis (LAMBDA), part of the
  High Energy Astrophysics Science Archive Center
  (HEASARC). HEASARC/LAMBDA is a service of the Astrophysics Science
  Division at the NASA Goddard Space Flight Center. This publication
  makes use of data products from the Wide-field Infrared Survey
  Explorer, which is a joint project of the University of California,
  Los Angeles, and the Jet Propulsion Laboratory/California Institute
  of Technology, funded by the National Aeronautics and Space
  Administration. This work has made use of data from the European
  Space Agency (ESA) mission {\it Gaia}
  (\url{https://www.cosmos.esa.int/gaia}), processed by the {\it Gaia}
  Data Processing and Analysis Consortium (DPAC,
  \url{https://www.cosmos.esa.int/web/gaia/dpac/consortium}). Funding
  for the DPAC has been provided by national institutions, in
  particular the institutions participating in the {\it Gaia}
  Multilateral Agreement.  We acknowledge the use of data provided by
  the Centre d'Analyse de Données Etendues (CADE), a service of
  IRAP-UPS/CNRS (http://cade.irap.omp.eu, \citealt{paradis:2012}).
  This paper and related research have been conducted during and with
  the support of the Italian national inter-university PhD programme
  in Space Science and Technology. Work on this article was produced
  while attending the PhD program in PhD in Space Science and
  Technology at the University of Trento, Cycle XXXIX, with the
  support of a scholarship financed by the Ministerial Decree no. 118
  of 2nd March 2023, based on the NRRP - funded by the European Union
  - NextGenerationEU - Mission 4 "Education and Research", Component 1
  "Enhancement of the offer of educational services: from nurseries to
  universities” - Investment 4.1 “Extension of the number of research
  doctorates and innovative doctorates for public administration and
  cultural heritage” - CUP E66E23000110001.
\end{acknowledgements}

%-------------------------------------------------------------
%                                       Table with references 
%-------------------------------------------------------------
%

\bibliographystyle{aa}
\bibliography{../../common/CG_bibliography,references,../../common/Planck_bib}
\end{document}